\documentclass[journal]{IEEEtran}
\usepackage[utf8]{inputenc}
\usepackage{mathtools}
\usepackage{algorithmic}
\usepackage{amssymb}
\usepackage[ruled]{algorithm}
\usepackage{theorem}
\usepackage[lowtilde]{url}
\usepackage{enumerate}
\usepackage{prettyref}
\usepackage{tikz}
\usetikzlibrary{calc,shapes,positioning}
\tikzset{varNode/.style={circle,draw,minimum size=4mm, inner sep=0.5mm}}
\tikzset{checkNode/.style={rectangle,draw,minimum size=4mm}}
\tikzset{>=stealth}
\tikzset{dot/.style={fill,circle,minimum size=1mm,inner sep=0mm}}
\tikzset{ffgNode/.style={inner sep=1mm,minimum size=3mm,draw}}

\DeclareMathOperator*{\argmin}{arg\,min}
\DeclareMathOperator*{\argmax}{arg\,max}
\DeclareMathOperator{\conv}{conv}
\DeclareMathOperator{\supp}{supp}
\DeclareMathOperator{\Aut}{Aut}
\newcommand{\F}{\mathbb{F}}
\newcommand{\abs}[1]{\lvert #1 \rvert}

\theoremstyle{plain}
\newtheorem{thm}{Theorem}[section]
\newtheorem{lem}[thm]{Lemma}
\newtheorem{defn}[thm]{Definition}

\usepackage{xspace}
\newcommand{\etal}{\textit{et al.}\xspace}
\newcommand{\ie}{i.\,e.\xspace}
\newcommand{\eg}{e.\,g.\xspace}
\newcommand{\tk}{\text{,}}

\newrefformat{fig}{Fig.~\ref{#1}}

\begin{document}
\title{Mathematical Programming Decoding of Binary Linear Codes: Theory and
Algorithms}
\author{
Michael Helmling,
Stefan Ruzika,
Ak\i{}n Tanatmis
\thanks{This work was supported in part by the
Center of Mathematical and Computational Modeling of the University of
Kaiserslautern.}%
\thanks{M.~Helmling, S.~Ruzika and A.~Tanatmis are with Department of Mathematics, University of Kaiserslautern, Erwin-Schroedinger-Strasse, 67663 Kaiserslautern, Germany. Email: \{helmling, ruzika\}@mathematik.uni-kl.de}
}

\maketitle

\begin{abstract}
Mathematical programming is a branch of applied mathematics and has recently
been used to derive new decoding approaches, challenging established but often
heuristic algorithms based on iterative message passing.
Concepts from mathematical programming used in the context of decoding include
linear, integer, and nonlinear programming, network flows, notions of duality as
well as matroid and polyhedral theory. This survey article reviews and
categorizes decoding methods based on mathematical programming approaches for
binary linear codes over binary-input memoryless symmetric channels.
\end{abstract}


\IEEEpeerreviewmaketitle

\section{Introduction} \label{intro}
Based on an integer programming (IP)\footnote{See the table on page \pageref{tab:abbr} for a
list of the acronyms used.} formulation of the maximum likelihood decoding
(MLD) problem for binary linear codes, linear programming decoding (LPD) was
introduced by Feldman \etal \cite{feld03,FeWaKa}. Since then, LPD has been
intensively studied in a variety of articles especially dealing with low-density
parity-check (LDPC) codes. LDPC codes are generally decoded by heuristic
approaches called iterative message passing decoding (IMPD) subsuming
sum-product algorithm decoding (SPAD) \cite{KsFrLoe,AjiMc} and min-sum
algorithm decoding (MSAD) \cite{wiberg}. In these algorithms, probabilistic
information is iteratively exchanged and updated between component decoders.
Initial messages are derived from the channel output. IMPD exploits the sparse
structure of parity-check matrices of LDPC and turbo codes very well and
achieves good performance. However, IMPD approaches are neither guaranteed to
converge nor do they have the maximum likelihood (ML) certificate property,
\ie, if the output is a codeword, it is not necessarily the ML codeword.
Furthermore, performance of IMPD is poor for arbitrary linear block codes with a
dense parity-check matrix.
In contrast, LPD offers some advantages and thus has become an important
alternative decoding technique. First, this approach is derived from the
discipline of mathematical programming which provides analytical statements on 
convergence, complexity, and correctness of decoding algorithms. Second, LPD is
not limited to sparse matrices. 

This article is organized as follows. In Section~\ref{notation}, notation is
fixed and well-known but relevant results from coding theory and polyhedral
theory are recalled. Complexity and polyhedral properties of MLD are discussed
in Section~\ref{CompPol}. In Section~\ref{basics} a general description of LPD
is given. Several linear programming (LP) formulations dedicated to codes with
low-density parity-check matrices, codes with high-density parity-check
matrices, and turbo-like codes are categorized and their commonalities and
differences are emphasized in Section~\ref{lprelaxations}. Based on these LP
formulations, different streams of research on LPD have evolved. Methods
focusing on efficient realization of LPD are summarized in Section~\ref{efflp},
while approaches improving the error-correcting performance of LPD at the cost
of increased complexity are reviewed in Section~\ref{incper}. Some concluding
comments are made in Section~\ref{concl}.

\section{Basics and Notation} \label{notation}
This section briefly introduces a number of definitions and results from
linear coding theory and polyhedral theory which are most fundamental for the
subsequent text.

A binary linear block code $C$ with cardinality $2^k$ and block length $n$ is a
$k$-dimensional subspace of the vector space $\{0,1\}^n$ defined over the binary
field $\F_2$. $C \subseteq \{0,1 \}^n$ is given by $k$ basis vectors of length
$n$ which are arranged in a $k \times n$ matrix $G$, called the generator matrix
of the code $C$.\footnote{Note that single vectors in this paper are generally
column vectors; however, in coding
theory they are often used as rows of matrices. The transposition of column
vector $a$ makes it a row vector, denoted by $a^T$.}    

The orthogonal subspace $C^\perp$ of $C$ is defined as
\[
C^\bot=\left\{y \in \{0,1\}^n : \sum_{j=1}^nx_jy_j \equiv  0\,
(\operatorname{mod} 2) \text{ for all } x \in C \right\}
\]
and has dimension $n-k$. It can also be interpreted as a binary linear code of
dimension $n-k$ which is referred to as the dual code of $C$. A matrix $H \in
\{0,1\}^{m \times n}$ whose $m\geq n-k$ rows form a spanning set of $C^\perp$
is called a parity-check matrix of $C$. It follows from this definition that $C$
is the null space of $H$ and thus a vector $x \in \{0,1\}^n$ is contained in $C$
if and only if $Hx \equiv 0 \pmod 2$. Normally, $m=n-k$ and the rows of $H \in
\{0,1\}^{(n-k)\times n}$ constitute a basis of $C^\perp$. It should be pointed
out, however, that most LPD approaches (see Section~\ref{incper}) benefit from
parity-check matrices being extended by redundant rows. Moreover, additional
rows of $H$ never degrade the error-correcting performance of LPD. This is a
major difference to IMPD which is generally weakened by redundant parity checks,
since they introduce cycles to the Tanner graph.

Let $x$, $x'$ $\in \{0,1\}^n$. The Hamming distance between $x$ and $x'$ is the
number of entries (bits) with different values, \ie, $d(x,x')=\left|\{1 \leq j
\leq n:\, x_j \neq x'_j \}\right|$. The minimum (Hamming) distance of a code,
$d(C)$, is given by $d(C)=\min\{d(x,x'):x,x'\in C, x \neq x'\}$. The Hamming
weight of a codeword $x \in C$ is defined as $w(x)=d(x,0)$, i.e., the number of
ones in $x$. The minimum Hamming weight of $C$ is $w(C)=\min\{w(x):x\in C, x
\neq 0\}$. For binary linear codes it holds that $d(C)=w(C)$. The
error-correcting performance of a code is, at least at high
signal-to-noise ratio (SNR), closely related to its minimum distance.

Let $A \in \mathbb{R}^{m \times n}$ denote an $m \times n$ matrix and
$I=\{1,\ldots,m\}$,  $J=\{1,\ldots,n\}$ be the row and column index sets of $A$,
respectively. The entry in row $i \in I$ and column $j \in J$ of $A$ is given by
$A_{i,j}$. The $i^{\text{th}}$ row and $j^{\text{th}}$ column of $A$ are denoted
by $A_{i,.}$ and $A_{.,j}$, respectively. A vector $e \in \mathbb{R}^m$ is
called the $i^\text{th}$ unit column vector if $e_i=1$, $i \in I$, and $e_h=0$
for all $h \in I \setminus \{i\}$. 

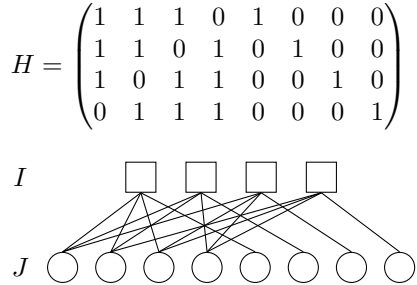
\begin{figure}[htp!]
 \centering
$H = \begin{pmatrix}1&1&1&0&1&0&0&0\\
 1&1&0&1&0&1&0&0\\
 1&0&1&1&0&0&1&0\\
 0&1&1&1&0&0&0&1\end{pmatrix}$\\[4mm]
\begin{tikzpicture}[xscale=.8]
 \foreach \i in {1,...,4} {
   \node[checkNode] (check\i) at (\i,0) {};
 }
 \foreach \j in {1,...,8} {
   \node[varNode] (variable\j) at (\j * .8-1.1, -1.2) {};
 }
 \foreach \i/\j in
{1/1,1/2,1/3,1/5,2/1,2/2,2/4,2/6,3/1,3/3,3/4,3/7,4/2,4/3,4/4,4/8} {
   \draw (check\i.south) -- (variable\j.north);
 }
 \node at (-1,0) {$I$};
 \node at (-1,-1.2) {$J$};
\end{tikzpicture}
\caption{Parity-check matrix and Tanner graph of an (8,4) code.}
\label{fig1new}
\end{figure}

A parity-check matrix $H$ can be represented by a bipartite graph $G=(V,E)$,
called its Tanner graph (Fig.~\ref{fig1new}). The vertex set $V$ of $G$ consists of the two disjoint
node sets $I$ and $J$. The nodes in $I$ are referred to as check nodes and
correspond to the rows of $H$ whereas the nodes in $J$ are referred to as
variable nodes and correspond to columns of $H$. An edge $[i,j] \in E$ connects
node $i$ and $j$ if and only if $H_{i,j}=1$.
Let $N_i = \{j \in J: H_{ij} = 1 \}$ denote the index set of variables
incident to check node $i$, and analogously $N_j = \{i \in I: H_{ij}=1\}$ for $j \in J$.
The degree of a check node $i$ is the number of edges incident to node $i$ in
the Tanner graph or, equivalently, $d_c(i)= \abs{N_i}$. The maximum
check node degree $d_c^{\max}$ is the degree of the check node $i \in I$ with
the largest number of incident edges. The degree of a variable node $j$,
$d_v(j)$, and the maximum variable node degree $d_v^{\max}$ are defined
analogously.

Tanner graphs are an example of factor graphs, a general concept of graphical models
which is prevalently used to describe probabilistic systems and related algorithms.
The term stems from viewing the graph as the representation of some global
function in several variables that factors into a product of subfunctions, each depending
only on a subset of the
variables. In case of Tanner graphs, the global function is the indicator function
of the code, and the subfunctions are the parity-checks according to single rows of $H$.
A different type of factor graphs will appear later in order to describe turbo codes. Far beyond these purely descriptive purpose,
factor graphs have proven successful in modern coding theory primarily in the
context of describing and analyzing IMPD algorithms. See \cite{loeligerFactor}
for a more elaborate introduction.

Let $C$ be a binary linear code with parity-check matrix $H$ and $x \in C
\subseteq \{0,1\}^n$. The index set $\supp(x)=\{j \in J:x_j=1\}$ is called the
support of the codeword $x$. A codeword $0 \neq x \in C$ is called a minimal
codeword if there is no codeword $0 \neq y \in C$ such that $\supp(y) \subseteq
\supp(x)$. Finally, $D$ is called a minor code of $C$ if $D$
can be obtained from $D$ by a series of shortening and puncturing operations.

The relationship between binary linear codes and polyhedral theory follows from
the observation that a binary linear code can be considered a set of points in
$\mathbb{R}^n$, \ie, $C \subseteq \{0,1 \}^n \subseteq \mathbb{R}^n$. In the
following, some relevant results from polyhedral theory are recalled. For a
comprehensive review on polyhedral theory the reader is referred to
\cite{NemWol}.
\begin{defn}
A subset $\mathcal{P}(A,b) \subseteq \mathbb{R}^n$ such that $\mathcal{P}(A,b) =
\{\nu \in \mathbb{R}^n: A\nu \leq b\}$ where $A \in \mathbb{R}^{m \times n}$ and
$b \in \mathbb{R}^m$ is called a polyhedron.
\end{defn}
In this article, polyhedra are assumed to be rational, \ie, the entries of $A$ and $b$
are taken from $\mathbb Q$.
The $i^\text{th}$ row vector of $A$ and the $i^\text{th}$ entry of $b$ together
define a closed halfspace $\{\nu \in \mathbb{R}^n: \; A_{i,.}\nu \leq b_i \}$.
In other words, a polyhedron is the intersection of a finite set of closed
halfspaces. A bounded polyhedron is called a polytope. It is known from
polyhedral theory that a polytope can equivalently be defined as the convex hull
of a finite set of points. In this work, the convex hull of a binary linear code
$C$ is denoted by $\conv(C)$ and referred to as the codeword polytope. 

Some characteristics of a polyhedron are its dimension, faces, and facets. To
define them, the notion of a valid inequality is needed. 
\begin{defn} \label{valineqdef}
An inequality $r^T\nu \leq t$, where $r \in \mathbb{R}^n$ and $t \in
\mathbb{R}$, is valid for a set $\mathcal{P}(A,b) \subseteq \mathbb{R}^n$ if
$\mathcal{P}(A,b) \subseteq \{\nu : r^T\nu \leq t\}$.
\end{defn}
The following definition of an active inequality is used in several LPD
algorithms.
\begin{defn}
An inequality  $r^T\nu\leq t$, where $r$, $\nu$ $\in \mathbb{R}^n$ and $t \in
\mathbb{R}$, is active at $\nu^* \in \mathbb{R}^n$ if $r^T \nu^* = t$.    
\end{defn}
Valid inequalities which contain points of $\mathcal{P}(A,b)$ are of special
interest.
\begin{defn} \label{face}
Let $\mathcal{P}(A,b) \subseteq \mathbb{R}^n$ be a polyhedron, let $r^T\nu \leq
t$ be a valid inequality for $\mathcal{P}(A,b)$ and define $F=\{\nu \in
\mathcal{P}(A,b): r^T\nu = t\}$. Then $F$ is called a face of
$\mathcal{P}(A,b)$. $F$ is a proper face if $F \neq \emptyset$ and $F \neq
\mathcal{P}(A,b)$. 
\end{defn}
The dimension $\dim(\mathcal{P}(A,b))$ of $\mathcal{P}(A,b) \subseteq
\mathbb{R}^n$ is given by the maximum number of affinely independent points in
$\mathcal{P}(A,b)$ minus one. Recall that a set of vectors
$v^1,\dotsc,v^k$ is affinely independent if the system $\{\sum_{i=1}^k \lambda_k
v^k = 0,\, \sum_{i=1}^k \lambda_k = 0\}$ has no solution other than
$\lambda_i=0$ for $i=1,\dotsc,k$. If $\dim(\mathcal{P}(A,b))=n$, then the
polyhedron is full-dimensional. It is a well-known result that if
$\mathcal{P}(A,b)$ is not full-dimensional, then there exists at least one
inequality $A_{i,.}\nu \leq b_i$ such that $A_{i,.}\nu = b_i$ holds for all $\nu
\in \mathcal{P}(A,b)$ (see \eg
\cite{NemWol}). Also, we have $\dim(F) \leq \dim(\mathcal{P}(A,b))-1$ for any
proper face of $\mathcal{P}(A,b)$. A face $F \neq \emptyset$ of
$\mathcal{P}(A,b)$ is called a facet of $\mathcal{P}(A,b)$ if $\dim(F)=
\dim(\mathcal{P}(A,b))-1$. 

In the set of inequalities defined by $(A,b)$, some inequalities $A_{i,.}\nu
\leq b_i$ may be redundant, \ie, dropping these inequalities does not change
the solution set defined by $A\nu \leq b$. A standard result
states that the facet-defining inequalities give a complete non-redundant
description of a polyhedron $\mathcal{P}(A,b)$ \cite{NemWol}.

A point $\nu \in \mathcal{P}(A,b)$ is called a vertex of $\mathcal{P}(A,b)$ if
there exist no two other points $\nu^1, \nu^2 \in \mathcal{P}(A,b)$ such that
$\nu=\mu_1\nu^1+\mu_2\nu^2$ with $0 \leq \mu_1 \leq 1$, $0 \leq \mu_2 \leq 1$,
and $\mu_1+\mu_2=1$. Alternatively, vertices are zero dimensional faces of
$\mathcal{P}(A,b)$. In an LP problem, a linear cost function is minimized on a
polyhedron, \ie, $\min \{ c^Tx:x \in \mathcal{P}(A,b) \}$, $c \in
\mathbb{R}^n$. Unless the LP is infeasible or unbounded, the minimum is attained
on one of the vertices. 

The number of constraints of an LP problem may be very large, \eg Section~\ref{lprelaxations} contains LPD formulations whose description
complexity grows exponentially with the block length for general
codes. In such a case it would be desirable to only include the constraints
which are necessary to determine the optimal solution of the LP with respect to
a given objective function. This can be accomplished by iteratively solving the
associated separation problem, defined as follows.
\begin{defn} \label{defn:SepAlg} Let $\mathcal{P}(A,b) \subset \mathbb{R}^n$ be
a rational polyhedron and $\nu^* \in \mathbb{R}^n$ a rational vector. The
separation problem is to either conclude that $\nu^* \in \mathcal{P}(A,b)$ or,
if not, find a rational vector $(r,t) \in \mathbb{R}^n \times \mathbb{R}$ such
that $r^T\nu \leq t$ for all $\nu \in \mathcal{P}(A,b)$ and $r^T\nu^* > t$.
In the latter case, $(r,t)$ is called a valid cut.
\end{defn}
We will see applications of this approach in Sections~\ref{efflp} and
\ref{incper}.

There is a famous result about the equivalence of optimization and separation by
Gr\"otschel \etal \cite{GrLoSch}.
\begin{thm}\label{OptSep} 
Let $\mathcal{P}$ be a proper class of polyhedra (see \eg \cite{NemWol} for a
definition). The optimization problem for $\mathcal{P}$ is polynomial time
solvable if and only if the separation problem is polynomial time solvable.
\end{thm}

\section{Complexity and Polyhedral Properties}\label{CompPol}
In this section, after referencing important NP-hardness results for the
decoding problem, we state useful properties of the codeword polytope,
exploiting a
close relation between coding and matroid theory.

Integer programming provides powerful means for modeling several real-world
problems. MLD for binary linear codes is modeled as an IP problem in
\cite{FeWaKa,breitbach}. Let $y \in \mathbb{R}^n$ be the channel output.
In MLD the probability (or, in case of a continuous-output channel, the
probability density) $P(y | x)$ is maximized over all codewords $x \in C$.
Let $x^*$ denote the ML codeword. It is shown in \cite{feld03} that for a
symmetric memoryless channel the calculation of $x^\ast$ amounts to the
minimization of a linear cost function, namely
\begin{equation}
x^*= \argmax_{x \in C} P(y | x) = \argmin_{x \in C} \sum_{j=1}^n \lambda_jx_j
\text{,} \label{eq:LinearCostFunction}
\end{equation}
where the values $\lambda_j = \textnormal{log}
\frac{P(y_j|x_j=0)}{P(y_j|x_j=1)}$ are the so-called log-likelihood ratios
(LLR). Consequently the IP formulation of MLD is implicitly given as
\begin{align}
\min \{ \lambda^Tx :x \in C \} \label{eq:MLdecoding}.
\end{align}
Berlekamp \etal have shown that MLD is NP-hard in \cite{BeMcTi} by a
polynomial-time reduction of the three-dimensional matching problem to the
decision version of MLD. An alternative proof is via
matroid theory: as shall be exposed shortly, there is a one-to-one correspondence
between binary matroids and binary linear codes. In virtue of this analogy, MLD
is equivalent to the minimum-weight
cycle problem on binary matroids. Since the latter contains the max-cut problem, which
is known to be NP-hard \cite{Karp72}, as a special case, the NP-hardness of MLD follows.

Another problem of interest in the framework of coding theory is the computation
of the minimum distance of a given code. Berlekamp \etal \cite{BeMcTi}
conjectured that computing the distance of a binary linear code is NP-hard as
well, which was proved by Vardy \cite{Vardy} about two decades later. The
minimum distance problem can again be reformulated in a matroid theoretic
setting. In 1969 Welsh \cite{Welsh69} formulated it as the problem of finding a
minimum cardinality circuit in linear matroids.

In the following, we assume $C \subseteq \{ 0 , 1 \}^n$ to be canonically
embedded in $\mathbb R^n$ when referring to $\conv(C)$ (see
\prettyref{fig:codepoly} for an example). Replacing $C$ by $\conv(C)$ in
(\ref{eq:MLdecoding}) leads to a linear programming problem over a polytope with
integer vertices. In general, computing an explicit representation of $\conv(C)$
is intractable. Nevertheless, some properties of $\conv(C)$ are known from
matroid theory due to the equivalence of binary linear codes and binary
matroids. In the following, some definitions and results from matroid theory are
presented. An extensive investigation of matroids can be found in \cite{Oxley}
or \cite{Welsh}.  The definition of a matroid in general is rather technical.
\begin{defn}
A matroid $\mathcal{M}$ is an ordered pair $\mathcal{M} = (J, \mathcal{U})$
where $J$ is a finite ground set and $\mathcal{U}$ is a collection of subsets of
$J$, called the independent sets, such that (a) -- (c) hold.
\begin{enumerate}[(a)]
 \item  $\emptyset \in \mathcal{U}$.
\item If $u \in \mathcal{U}$ and $v \subset u$, then $v \in \mathcal{U}$.
\item If $u_1, u_2 \in \mathcal{U}$ and $\left|u_1\right| < \left|u_2\right|$
then there exists $j \in u_2 \setminus u_1$ such that $u_1 \cup \{ j \} \in
\mathcal{U}$.
\end{enumerate}
\end{defn}
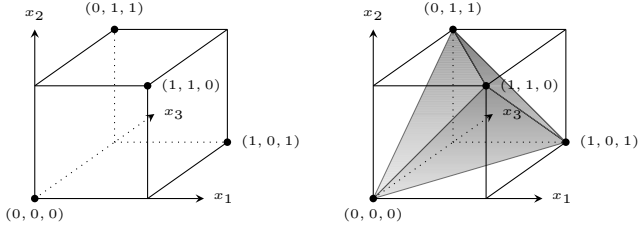
\begin{figure}

\begin{tikzpicture}[x={(1cm,0cm)},y={(0cm,1cm)},z={(0.707cm,0.5cm)},scale=1.5,font=\tiny]
\def\drawcube{
  \draw[->] (0,0,0) -- (1.5,0,0) node[right] {$x_1$};
  \draw[->] (0,0,0) -- (0,1.5,0) node[above] {$x_2$};
  \draw[->,dotted] (0,0,0) -- (0,0,1.5) node[right] {$x_3$};
  \draw[dotted] (0,0,1) -- (1,0,1);
  \draw[dotted] (0,0,1) -- (0,1,1);
  \draw (1,0,0) -- (1,0,1);
  \draw (0,1,1) -- (0,1,0);
  \draw (0,1,1) -- (1,1,1);
  \draw (1,0,1) -- (1,1,1);
  \draw (1,0,0) -- (1,1,0);
  \draw (0,1,0) -- (1,1,0);
  \draw (1,1,0) -- (1,1,1);
  \node[dot] at (0,0,0) {} node[below] {$(0,0,0)$};
  \node[dot,label=right:{$(1,1,0)$}] at (1,1,0) {};
  \node[dot,label=right:{$(1,0,1)$}] at (1,0,1) {};
  \node[dot,label=above:{$(0,1,1)$}] at (0,1,1) {};
}
\drawcube
\begin{scope}[xshift=3cm]
  \draw[fill=black!50,shade,opacity=.5] (0,0,0) -- (1,0,1) -- (1,1,0) -- cycle;
  \draw[fill=black!30,shade,opacity=.5] (0,0,0) -- (1,1,0) -- (0,1,1) -- cycle;
  \draw[fill,shade,opacity=.5] (0,1,1) -- (1,1,0) -- (1,0,1) -- cycle;
  \drawcube
\end{scope}
 \end{tikzpicture}
\caption{The codewords of the single parity-check code $C=\{x \in \F_2^3:
x_1+x_2+x_3 \equiv 0 \pmod 2\}$ and the polytope $\conv(C)$ in $\mathbb
R^3$.}\label{fig:codepoly}
\end{figure}

In this work, the class of $\mathbb{F}_2$-representable (\ie, binary) matroids
is of interest. A binary $m \times n$ matrix $H$ defines an
$\mathbb{F}_2$-representable matroid $\mathcal{M} \left[ H \right]$ as follows.
The ground set $ J = \{ 1, \ldots, n \}$ is defined to be the index set of the
columns of $H$. A subset $U \subseteq J$ is independent if and only if the
column vectors $H_{ .,u }$, $u \in U$ are linearly independent in the vector
space defined over the field $\mathbb{F}_2$. A minimal dependent set, \ie, a
set $\mathcal{V} \in 2^{ J } \setminus \mathcal{U}$ such that all proper subsets
of $\mathcal{V}$ are in $\mathcal{U}$, is called a circuit of $\mathcal{M}
\left[ H \right]$. If a subset of $J$ is a disjoint union of circuits then it is
called a cycle.

The incidence vector $x^{\mathcal{C}} \in \mathbb{R}^n$ corresponding to a cycle
$\mathcal{C} \subseteq J$ is defined by
\[ x_j^{\mathcal{C}} = \begin{cases}
                         1 & \text{if }j\in C,\\
                         0 & \text{if }j \notin C.
                        \end{cases}\]
The cycle polytope is the convex hull of the incidence vectors corresponding to
all cycles of a binary matroid. 

Some more relationships between coding theory and matroid theory (see also
\cite{Kashyap}) can be listed: a binary linear code corresponds to a binary
matroid, the support of a codeword corresponds to a cycle
(therefore, each codeword corresponds to the incidence vector of a cycle), the
support of a minimal codeword corresponds to a circuit, and the codeword
polytope $\conv(C)$ corresponds to the cycle polytope. Let $H$ be a binary
matrix, $\mathcal{M} \left[ H \right]$ be the binary matroid defined by $H$ ($H$
is a representation matrix of $\mathcal{M} \left[ H \right]$) and $C$ be the
binary linear code defined by $H$ ($H$ is a parity-check matrix of $C$). It can
easily be shown that the dual $C^\bot$ of $C$ is the same object as the dual of
the binary matroid $\mathcal{M} \left[ H \right]$. We denote the dual matroid by
$\mathcal{M} \left[ G \right]$, where $G$ is the generator matrix of $C$.
Usually the matroid related terms are dualized by the prefix “co”. For example,
the circuits and cycles of a dual matroid are called cocircuits and cocycles,
respectively. The supports of minimal codewords and the supports of codewords in
$C^\bot$ are associated with cocircuits and cocycles of $\mathcal{M} \left[ H
\right]$, respectively.

A minor of a parent matroid $\mathcal{M}=(J, \mathcal{U})$ is the sub-matroid
obtained from $\mathcal{M}$ after any combination of contraction and restriction
operations (see \eg \cite{Oxley}). In the context of coding theory, contraction
corresponds to puncturing, \ie, the deletion of one or more columns from the
generator matrix of a parent code, and restriction corresponds to shortening,
\ie, the deletion of one or more columns from the parity-check matrix of a parent
code. 

Next, some results from Barahona and Grötschel \cite{BaraGr} which are related
to the structure of the cycle polytope are rewritten in terms of coding theory.
Kashyap provides a similar transfer in \cite{Kashyap06}. Several results are
collected in Theorem~\ref{cyclepolytoperesults}.
\begin{thm} \label{cyclepolytoperesults}
Let $C$ be a binary linear code.
\begin{enumerate}[(a)]
\item If $d(C^{\bot}) \geq 3$  then the codeword polytope is full-dimensional.
\item\label{item:test} The box inequalities
\begin{equation}
0 \leq x_j \leq 1 \quad\text{for all } j \in J \label{trivi}
\end{equation}
and the cocircuit inequalities
\begin{equation}
\begin{multlined}
\sum_{j \in \mathcal{F}} x_j - \sum_{j \in \supp(q) \setminus \mathcal{F}} x_j
\leq \left| \mathcal{F} \right| - 1 \\
\text{ for all } \mathcal{F} \subseteq \supp(q) \text{ with } \left| \mathcal{F}
\right| \text{ odd}, 
\end{multlined}\label{cfineq}\end{equation}
where $\supp(q)$ is the support of a dual minimal codeword $q$, are valid for
the codeword polytope.
\item The box inequalities $x_j \geq 0$, $x_j \leq 1$ define facets of the
codeword polytope if $d(C^{\bot}) \geq 3$ and $j \in J$ is not contained in the
support of a codeword in $C^\bot$ with weight three.
\item If $d(C^{\bot}) \geq 3$ and $C$ does not contain $H_7^{\bot}$ ((7,3,4)
simplex code) as a minor, and if there exists a dual minimal codeword $q$ of
weight $3$, then the cocircuit inequalities derived from $\supp(q)$ are facets
of $\conv(C)$.
\end{enumerate}
\end{thm}
Part~(\ref{item:test}) of Theorem~\ref{cyclepolytoperesults} implies that the
set of cocircuit inequalities derived from the supports of all dual minimal
codewords provide a relaxation of the codeword polytope. In the polyhedral
analysis of the codeword polytope the symmetry property stated below plays an
important role.
\begin{thm} \label{bathe}
\cite{BaraGr} If $a^Tx \leq \alpha$ defines a face of $\conv(C)$ of dimension
$d$, and $y$ is a codeword, then the inequality $\bar{a}^T x \leq \bar{\alpha}$
also defines a face of $\conv(C)$ of dimension $d$, where
\begin{eqnarray*}
\bar{a}_j &=&\left\{
\begin{array}{rl}
a_j & \text{ if } j \notin \supp(y)\text{,} \\
-a_j & \text{ if } j \in \supp(y)\text{,} \\
\end{array} \right.
\end{eqnarray*}
and $\bar{\alpha}=\alpha - a^Ty$.
\end{thm}

Using this theorem, a complete description of $\conv(C)$ can be derived from all
facets containing a single codeword \cite{BaraGr}.

Let $q$ be a dual minimal codeword. To identify if the cocircuit inequalities
derived from $\supp(q)$ are facet-defining it should be checked if $\supp(q)$
has a chord. For the formal definition of chord, the
symmetric difference $\triangle$ which operates on two finite sets is used,
defined by 
$A \triangle B = (A \setminus B) \cup (B \setminus A)$. Note that if
$A=\supp(q_1)$, $B=\supp(q_2)$ and $\supp(q_0) = A\triangle B$, then $q_0 \equiv
q_1 + q_2 \pmod 2$.
   
\begin{defn}
Let $q_0,q_1,q_2 \in C^\bot$ be dual minimal codewords. If $\supp(q_0) =
\supp(q_1) \triangle \supp(q_2)$ and $\supp(q_1) \cap \supp(q_2) =\{j\}$, then
$j$ is called a chord of $\supp(q_0)$.
\end{defn}

\begin{thm}\label{facet2}
\cite{BaraGr} Let $C$ be a binary linear code without the $(7,3,4)$ simplex code as
a minor and let $\supp(q)$ be the support of a dual minimal codeword with
Hamming
weight at least $3$ and without chord. Then 
for all $\mathcal F \subseteq \supp(q)$ with $\abs{\mathcal F}$ odd, the
inequality
\[
\sum_{j \in \mathcal{F}}x_j - \sum_{j \in \supp(q) \setminus \mathcal{F}} x_j
\leq \left|\mathcal{F}\right|-1
\]
defines a facet of $\conv(C)$. 
\end{thm}

Optimizing a linear cost function over the cycle polytope, known as the cycle
problem in terms of matroid theory, is investigated by Gr\"otschel and Truemper
\cite{GrTr1}. The work of Feldman \etal \cite{FeWaKa} enables to use the matroid
theoretic results in the coding theory context.
As shown above, solving the MLD problem for a binary linear code is equivalent
to solving the cycle problem on a binary matroid.
In \cite{GrTr1}, binary matroids for which the cycle problem can be solved in
polynomial time are classified, based on Seymour's matroid decomposition theory
\cite{Seymour}. Kashyap \cite{Kashyap} shows that results from \cite{GrTr1} are
directly applicable to binary linear codes. The MLD problem as well as the
minimum distance problem can be solved in polynomial time for the code families
for which the cycle problem on the associated binary matroid can be solved in
polynomial time. This code family is called polynomially almost-graphic codes
\cite{Kashyap}.

An interesting subclass of polynomially almost-graphic codes are geometrically
perfect codes. Kashyap translates the sum of circuits  property (see
\cite{GrTr1}) to the realm of binary linear codes. If the binary matroid
associated with code $C$ has the sum of circuits property then  $\conv(C)$ can
be described completely and non-redundantly by the box inequalities
\eqref{trivi} and the cocircuit inequalities \eqref{cfineq}. These codes are
referred to as geometrically perfect codes in \cite{Kashyap}. The associated
binary matroids of geometrically perfect codes can be decomposed in polynomial
time into its minors which are either graphic (see \cite{Oxley}) or contained in
a finite list of matroids. 

From a coding theoretic point of view, a family of error-correcting codes is
asymptotically bad if either dimension or minimum distance grows only
sublinearly with the code length. Kashyap proves that the
family of geometrically perfect codes unfortunately fulfills this property. We
refer to \cite{Kashyap} for the generalizations of this result. 

\section{Basics of LPD} \label{basics}
LPD was first introduced in \cite{FeWaKa}. This decoding method is, in
principle, applicable to any binary linear code over any binary input memoryless
channel.\footnote{
In fact, Flanagan \etal \cite{FlanaganNonBinary} have recently generalized a
substantial portion of the LPD theory to the nonbinary case. Similarly, work has
been done to include channels with memory; see \eg
\cite{Cohen08Polya}.} In this section, we review the basics of the LPD approach
based on \cite{feld03}. 

Although several structural properties of $\conv(C)$ are known, it is in general
infeasible to compute a concise description of $\conv(C)$ by means of linear
inequalities. In LPD, the linear cost function of the IP formulation is
minimized on a relaxed polytope $\mathcal{P}$ where $\conv(C)$ $\subseteq$
$\mathcal{P} \subseteq \mathbb{R}^n$. Such a relaxed polytope $\mathcal{P}$
should have the following desirable properties:
\begin{itemize}
 \item $\mathcal P$ should be easy to describe, and
 \item integral vertices of $\mathcal{P}$ should correspond to codewords.
\end{itemize}
  Together with the linear representation
\eqref{eq:LinearCostFunction} of the
likelihood function, this leads to one of the major benefits of LPD, the
so-called ML certificate property: If the LP decoder outputs an integral optimal
solution, it is guaranteed to be the ML codeword.
This is a remarkable difference to IMPD:
If no general optimality condition applies (see \eg \cite[Sec.\ 10.3]{Shulin}), there is no method to provably decide the optimality of a solution obtained by IMPD.

Each row (check node) $i \in I$ of a parity-check matrix $H$ defines the local
code 
\[C_i = \left\{x \in \{0,1\}^n: \sum_{j=1}^n H_{ij} x_j \equiv 0 \pmod
2\right\}\]
that consists of the bit sequences which satisfy the $i^\text{th}$ parity-check
constraint; these are called local codewords. A particularly interesting
relaxation of $\conv(C)$ is \[\mathcal{P}=\conv(C_1)  \cap  \dotsm \cap
\conv(C_m) \subseteq [0,1]^n\text{,}\] known as the fundamental polytope
\cite{VoKoGraph}. The vertices of the fundamental polytope, the so-called
pseudocodewords, are a superset of $C$, where the difference consists only
of non-integral vertices. Consequently, optimizing over $\mathcal P$ implies the
ML certificate property. These observations are formally stated in the following
result (note that $C=C_1 \cap \cdots \cap C_m$). 

\begin{lem}   \label{lem:lemfundpoly}
\cite{VoKoGraph} Let $\mathcal{P}=\conv(C_1)  \cap  \cdots \cap \conv(C_m)$. If
$C=C_1  \cap  \cdots  \cap C_m$ then $\conv(C) \subseteq \mathcal{P}$ and $C =
\mathcal P \cap \{0,1\}^n$.
\end{lem}
The description complexity of the convex hull of any local code $\conv(C_i)$ and
thus $\mathcal{P}$ is usually much smaller than the description complexity of
the codeword polytope $\conv(C)$. 

LPD can be written as optimizing the linear objective function on the
fundamental polytope $\mathcal P$, \ie,
\begin{align}
 \min \{ \lambda^Tx :x \in \mathcal{P} \}  \label{eq:LPdecoding}.
\end{align}

Based on (\ref{eq:LPdecoding}), the LPD algorithm which we refer to as bare
linear programming decoding (BLPD) is derived.
\begin{algorithm}
\textbf {Bare LP decoding (BLPD)}\hrule\vspace{1mm}
\textbf{Input:} $\lambda \in \mathbb{R}^n$, $\mathcal{P} \subseteq \left[ 0,1
\right]^n$.\\
\textbf{Output:}  ML codeword or \textsc{error}.

\begin{algorithmic}[1]
\STATE solve the LP given in (\ref{eq:LPdecoding})
\IF{LP solution $x^*$ is integral}
\STATE output $x^*$
\ELSE
\STATE output \textsc{error}
\ENDIF
\end{algorithmic}
\end{algorithm}

Because of the ML certificate property, if BLPD outputs a codeword, then it is
the ML codeword.
BLPD succeeds if the transmitted codeword is the unique optimum of the LP given
in (\ref{eq:LPdecoding}). BLPD fails if the optimal solution is non-integral or
the ML codeword is not the same as the transmitted codeword. Note that the
difference between the performance of BLPD and MLD is caused by the decoding
failures for which BLPD finds a non-integral optimal solution. It should be
emphasized that in case of multiple optima it is assumed that BLPD fails.

In some special cases, the fundamental polytope $\mathcal{P}$ is equivalent to
$\conv(C)$, \eg, if the underlying Tanner graph is a tree or forest
\cite{VoKoGraph}. In these cases MLD can be achieved by BLPD. Note that in those
cases also MSAD achieves MLD performance \cite{wiberg}.

Observe that the minimum distance of a code can be understood as the minimum
$\ell_1$ distance between any two different codewords of $C$. Likewise the
fractional distance
of the fundamental polytope $\mathcal{P}$ can be defined as follows.
\begin{defn}\cite{FeWaKa}
Let $V(\mathcal{P})$ be the set of vertices (pseudocodewords) of $\mathcal{P}$.
The fractional distance $d_{\text{frac}}(\mathcal{P})$ is the minimum $\ell_1$
distance between a codeword and any other vertex of $V(\mathcal{P})$, \ie
\begin{displaymath}
d_{\text{frac}}(\mathcal{P})= \min \left\{ \sum_{j=1}^n\left|x_j-v_j\right|: x
\in C, \; v \in V(\mathcal{P}), \; x \neq v \right\}\text{.}
\end{displaymath}
\end{defn}

It follows that the fractional distance is a lower bound for the minimum
distance of a code: $d(C) \geq d_{\text{frac}}(\mathcal{P})$. Moreover, 
both definitions are related as follows. Recall that on the binary symmetric
channel (BSC), MLD corrects at least $\left\lceil d(C)/2\right\rceil-1$ bit
flips. As shown in \cite{feld03}, LPD succeeds if at most $\lceil
d_\text{frac}(\mathcal P)/2\rceil -1$ errors occur on the BSC. 

Analogously to the minimum distance, the fractional distance is equivalent to
the minimum $\ell_1$ weight of a non-zero vertex of $\mathcal{P}$. This property
is used by the fractional
distance algorithm (FDA) to compute the fractional distance of a binary linear
code \cite{feld03}.
If $\mathcal M$ is the set of inequalities describing $\mathcal P$, let
$\mathcal M_I$ be the subset of those inequalities which are not active at the
all-zero codeword. Note that these are exactly
the inequalities with a non-zero right hand side. In FDA the weight function
$\sum_{j \in J}{x_j}$ is subsequently minimized on $\mathcal{P} \cap f$ for all
$f \in  \mathcal{M}_I$ in order to find the minimum-weight non-zero vertex of
$\mathcal{P}$.
\begin{algorithm}
\textbf{Fractional distance algorithm (FDA)}\hrule\vspace{1mm}
\textbf{Input:} $\mathcal{P} \subseteq \left[ 0,1 \right]^n$. \\
\textbf{Output:} Minimum-weight non-zero vertex of $\mathcal{P}$.

\begin{algorithmic}[1]
\FORALL{$f \in \mathcal{M}_I$}
\STATE Set $\mathcal{P}'=\mathcal{P} \cap f$.
\STATE Solve $\min \left\{ \sum_{j \in J}{x_j} :x \in \mathcal{P}' \right\}$.
\ENDFOR
\STATE Choose the minimum value obtained over all $\mathcal{P}'$.
\end{algorithmic}
\end{algorithm}

A more siginficant distance measure than $d_{\text{frac}}$ is the so-called
pseudo-distance which quantifies the probability that the optimal solution
under LPD changes from one vertex of $\mathcal P$ to another \cite{Forney+2001,VoKoGraph}.
Likewise, the minimum
pseudo-weight is defined as the minimum pseudo-distance from $0$ to any other
vertex of $\mathcal P$ and therefor identifies the vertex (pseudocodeword)
which is most likely to cause a decoding failure. Note that the pseudo-distance
takes the channel's probability measure into account and thus depends on the chosen
channel model.

Albeit no efficient algorithms are known to compute the exact minimum
pseudo-weight of the fundamental polytope of a code, promising heuristics 
as well as analytical bounds have been proposed \cite{VoKoGraph, Forney+2001, ChertkovPolytope}.

\section{LPD Formulations for Various Code Classes}\label{lprelaxations}
This section reviews various formulations of the polytope $\mathcal P$ from
\eqref{eq:LPdecoding}, leading to optimized versions of the general BLPD
algorithm for different classes of codes.

In Step~1 of BLPD the LP problem is solved by a general purpose LP solver. These
solvers usually employ the simplex method  since it performs well in practice.
The simplex method iteratively examines vertices of the underlying polytope
until the vertex corresponding to the optimal solution is reached. If there
exists a neighboring vertex for which the objective function can be improved in
the current step, the simplex method moves to this vertex. Otherwise it stops.
The procedure of moving from one vertex to an other is called a simplex
iteration. Details on the simplex algorithm can be found in classical books
about linear programming (see \eg \cite{Schrijver}).

The efficiency of the simplex method depends on the complexity of the constraint
set describing the underlying polytope. Several such explicit descriptions of
the fundamental polytope $\mathcal{P}$ have been proposed in the LPD literature.
Some can be used for any binary linear code whereas others are specialized for a
specific code class. Using alternative descriptions of $\mathcal{P}$,
alternative LP decoders are obtained.
In the following, we are going to present different LP formulations. 

\subsection{LP formulations for LDPC codes}
The solution algorithm referred to as BLPD in Section \ref{basics} was
introduced by Feldman \etal \cite{FeWaKa}. In order to describe $\mathcal{P}$
explicitly, three alternative constraint sets are suggested by the authors by
the formulations BLPD1, BLPD2, and BLPD3. In the following, some abbreviations
are used to denote both the formulation and the associated solution (decoding)
algorithm, \eg, solving an LP, subgradient optimization, neighborhood search.
The meaning will be clear from the context.

The first LP formulation, BLPD1, of \cite{FeWaKa} is applicable to LDPC codes. 
	\begin{alignat}{2}
	 & \min \lambda^Tx  \quad \quad {\text{(BLPD1)}} \nonumber\\
	 \text{s.t. } & \sum_{S \in E_i} w_{i,S} = 1 && i=1, \ldots, m
\label{LPD1con3}\\
	 & x_j = \sum_{\mathclap{\substack{S \in E_i \\\text{with } j \in S}}} w_{i,S}
&& \forall j \in N_i,\, i=1, \ldots, m \label{LPD1con4}\\ 
	 & 0 \leq x_j \leq 1 &&  j=1, \ldots, n \nonumber\\
	 & 0 \leq w_{i,S} \leq 1 && \forall S \in E_i,\,i=1, \ldots, m \nonumber
	\end{alignat}

Here, $E_i = \{S \subseteq N_i: \abs S \text{ even}\}$
is the set of valid bit
configurations within $N_i$. The auxiliary variables $w_{i,S}$ used in this
formulation indicate which bit configuration $S \in E_i$ is taken at parity
check $i$. In case of an integral solution,
\eqref{LPD1con3} ensures that exactly one such configuration is attained at
every checknode, while \eqref{LPD1con4} connects the actual code bits, modeled
by the variables $x_j$, to the auxiliary variables:
$x_j=1$ if and only if the set $S \in E_i$ contains $j$ for every check node
$i$. Note that here we consider the LP relaxation, so it is not guaranteed that
a solution of the above program is indeed integral.

A second linear programming formulation for LDPC codes, BLPD2, is obtained by
employing the so-called forbidden set (FS) inequalities \cite{DiGoWa}. The FS
inequalities are motivated by the observation that one can explicitly forbid
those value assignments to variables where $\abs{S}$ is odd. For all local
codewords in $C_i$ it holds that
	\begin{align*}
	&\sum_{j \in S} x_j - \sum_{j \in N_i \setminus S} x_j \leq \left| S \right| -
1 &&\forall S \in \Sigma_i
	\end{align*}
where $\Sigma_i=\left\{S \subseteq N_i : \left| S \right| \; \textnormal{odd}
\right\}$. Feldman \etal show in \cite{FeWaKa} that for each single parity-check
code $C_i$, the FS inequalities together with
the box inequalities $0 \leq x_j \leq 1$, $j \in J$
completely and non-redundantly describe $\conv(C_i)$ (the case
$\abs{N_i}=3$ as depicted in \prettyref{fig:codepoly} is the only exception where the box inequalities are not needed). In a more general setting,
Gr\"otschel proved this result for the cardinality homogeneous set systems
\cite{GroetCard}. 

If the rows of $H$ are considered as dual codewords, the set of FS inequalities
is a reinvention of cocircuit inequalities explained in Section \ref{CompPol}.
BLPD2 is given below.   
	\begin{align*}
	&  \min \lambda^Tx \quad\quad\text{(BLPD2)} \\
	 \text{s.t. }&\sum_{j \in S} x_j - \sum_{\mathclap{j \in N_i \setminus S}} x_j
\leq  \left| S \right| - 1 && \forall  S \in \Sigma_i , \; i=1, \ldots ,m \\
	& 0 \leq x_j \leq 1 && j=1, \ldots, n
	\end{align*}

Feldman \etal \cite{FeWaKa} apply BLPD using formulations BLPD1 or BLPD2 to LDPC
codes. Under the BSC, the error-correcting performance of BLPD is compared with
the MSAD on an random rate-$\frac{1}{2}$ LDPC code with $n=200$, $d_v=3$,
$d_c=6$; with the MSAD, SPAD on the random rate-$\frac{1}{4}$ LDPC code with
$n=200$, $d_v=3$, $d_c=4$; with the MSAD, SPAD, MLD on the random
rate-$\frac{1}{4}$ LDPC code with $n=60$, $d_v=3$, $d_c=4$. On these codes, BLPD
performs better than MSAD but worse than SPAD. Using BLPD2, the FDA is applied
to random rate-$\frac{1}{4}$ LDPC codes with $n=100,200,300,400$, $d_v=3$ and
$d_c=4$ from an ensemble of Gallager \cite{Gallager}. For $(n-1,n)$ Reed-Muller
codes \cite{Forney} with $4 \leq n \leq 512$ they compare the classical distance
with the fractional distance. The numerical results suggest that the gap between
both distances grows with increasing block length.

Another formulation for LDPC codes is given in
Section~\ref{subsec:messagePassing} in the context of efficient implementations.

In a remarkable work, Feldman and Stein \cite{FeldmanStein} have shown that the
Shannon capacity of a channel can be achieved with LP decoding, which implies
a polynomial-time decoder and the availability of an ML certificate. To this
end, they use a slightly modified version of BLPD1 restricted to expander codes,
which are a subclass of LDPC codes. See \cite{FeldmanStein} for a formal
definition of expander codes as well as the details of the corresponding
decoder.

\subsection{LP formulations for codes with high-density parity-check matrices}
The number of variables and constraints in BLPD1 as well as the number of
constraints in BLPD2 increase exponentially in the check node degree. Thus, for
codes with high-density parity-check matrices, BLPD1 and BLPD2 are
computationally inefficient. A polynomial-sized formulation, BLPD3, is based on
the parity polytope of Yannakakis \cite{Yannakakis}. There are two types of
auxiliary variables in BLPD3. The variable $p_{i,k}$ is set to one if $k$
variable nodes are set to one in the neighborhood of parity-check~$i$, for $k$
in the index set $K_i=\left\{0,2,\ldots,2\left\lfloor \frac{\abs{N_i}}{2}
\right\rfloor\right\}$. Furthermore, the variable $q_{j,i,k}$ is set to one if
variable node $j$ is one of the $k$ variable nodes set to one in the
neighborhood of check node $i$.
 	\begin{align*}
	 &  \min \lambda^Tx \quad \quad \text{(BLPD3)}\\
	 \text{s.t. }& x_j=\sum_{k \in K_i}q_{j,i,k} &&  i \in N_j,\, j=1, \ldots, n\\
	 & \sum_{k \in K_i}p_{i,k}=1 && i=1, \ldots m\\
	 & \sum_{j \in N_i}q_{j,i,k}=kp_{i,k} && k \in K_i,\,i=1, \ldots m \\
	 & 0 \leq x_j \leq 1 && j=1, \ldots, n\\
	 & 0 \leq p_{i,k} \leq 1 && k \in K_i,\, i=1, \ldots, m \\	
	 & 0 \leq q_{j,i,k} \leq p_{i,k} && k \in K_i, \, j=1, \ldots, n, i \in N_j	 	
	\end{align*}

Feldman \etal \cite{FeWaKa} show that BLPD1, BLPD2, and BLPD3 are equivalent in
the sense that the $x$-variables of the optimal solutions in all three
formulations take the same values.

\begin{figure}[htp!]
 \centering
\begin{tikzpicture}[node distance=.2cm and .5cm]
 \node[checkNode] (c) at (0,0) {+};
 \node[varNode,above left=of c] (v1) {$\nu_1$} edge (c);
 \node[varNode,above right=of c] (v3) {$\nu_3$} edge (c);
 \node[varNode,below left=of c] (v2) {$\nu_2$} edge (c);
 \node[varNode,below right=of c] (v4) {$\nu_4$} edge (c);
 \node[single arrow,minimum height=1cm,shape border uses incircle,fill=gray] at
(2,0) {};
 \begin{scope}[xshift=5cm]
  \node[varNode] (v5) at (0,0) {$\nu_5$};
  \node[checkNode,left=of v5] (c1) {+} edge (v5);
  \node[checkNode,right=of v5] (c2) {+} edge (v5);
  \node[varNode,above left=of c1] (v1) {$\nu_1$} edge (c1);
  \node[varNode,below left=of c1] (v2) {$\nu_2$} edge (c1);
  \node[varNode,above right=of c2] (v3) {$\nu_3$} edge (c2);
  \node[varNode,below right=of c2] (v4) {$\nu_4$} edge (c2);
 \end{scope}
\end{tikzpicture}

\caption{Check node decomposition.} 
\label{fig:decomp}
\vspace{-0mm}
\end{figure}
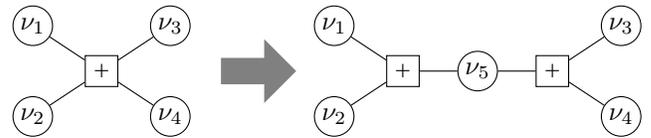

The number of variables and constraints in BLPD3 increases as $O(n^3)$. By
applying a decomposition approach, Yang \etal \cite{YaWaFe} show that an
alternative LP formulation which has size linear in the length and check node
degrees can be obtained (it should be noted that independently from
\cite{YaWaFe} a similar decomposition approach was also proposed in
\cite{ChertStep}). In the LP formulation of \cite{YaWaFe} a high degree check
node is decomposed into several low degree check nodes. Thus, the resulting
Tanner graph contains auxiliary check and variable nodes. \prettyref{fig:decomp}
illustrates this decomposition technique: a check node with degree $4$ is
decomposed into $2$ parity checks each with degree at most $3$. 
The parity-check nodes are illustrated by squares. In the example, original
variables are denoted by $\nu_1, \ldots, \nu_4$ while the auxiliary variable
node is named $\nu_5$. In general, this decomposition technique is iteratively
applied until every check node has degree less than $4$. The authors show that 
the total number of variables in the formulation is less than doubled by the
decomposition. For the details of the decomposition \cite{YaWaFe} is referred. 

For the ease of notation, suppose $K$ is the set of parity-check nodes after
decomposition. If $d_c(k)=3$, $k \in K$, then the parity-check constraint $k$ is
of the form $\nu^k_1 + \nu^k_2 + \nu^k_3 \equiv 0 \pmod 2$. Note that with our
notation some of these variables $\nu^k_s$ might represent the same variable
node $\nu_j$, e.g. $\nu_5$ from \prettyref{fig:decomp} would
appear in two constraints of the above form, as $\nu^1_s$ and $\nu^2_{s'}$,
respectively. Yang \etal show that the parity-check constraint $\nu^k_1 +
\nu^k_2 + \nu^k_3 \equiv 0 \pmod 2$ can be replaced by the linear constraints
$\nu^k_1 + \nu^k_2 + \nu^k_3 \leq 2, \nu^k_1 - \nu^k_2 - \nu^k_3 \leq 0, \nu^k_2
- \nu^k_1 - \nu^k_3 \leq 0, \nu^k_3 - \nu^k_1 - \nu^k_2 \leq 0$ (for a single
check node of degree $3$ the box inequalities are not needed). If $d_c(k)=2$
then $\nu^k_1=\nu^k_2$ along with the box constraints models the parity-check. 
The constraint set of the resulting LP formulation, which we call cascaded
linear programming decoding (CLPD), is the union of all constraints modeling the
$\left|K\right|$ parity checks.
\begin{align*}
&  \min \bar{\lambda}^T\nu \qquad \qquad \text{(CLPD)} \nonumber\\
\text{s.t. }&  \sum_{j \in S}  \nu^k_j - \sum_{\mathclap{j \in N_k \setminus S}}
\nu^k_j \leq  \left| S \right| - 1
   &\forall  S \in \Sigma_k , \; k=1, \ldots ,\left|K\right|\\
   &0 \leq \nu_j \leq 1&\text{if }d_c(i) \leq 2\; \forall\, i: j \in N_i
\end{align*}
In the objective function only the $\nu$ variables corresponding to the original
$x$ variables have non-zero coefficients. Thus, the objective function of CLPD
is the same as of BLPD1. The constraints in CLPD are the FS inequalities used in
BLPD2 with the property that the degree of the check node is less than $4$.

Yang \etal prove that the formulations introduced in \cite{FeWaKa} and CLPD are
equivalent. Again, equivalence is used in the sense that in an optimal solution,
the $x$-variables of BLPD1, BLPD2, BLPD3,
and the variables of the CLPD formulation which correspond to original
$x$-variables take the same values. Moreover, it is shown that CLPD can be used
in FDA. 
As a result, the computation of the fractional distance for codes with
high-density parity-check matrices is also facilitated. Note that using BLPD2,
the FDA algorithm has polynomial running time only for LDPC codes. If
$\mathcal{P}$ is described by the constraint set of CLPD, then in the first step
of the FDA, it is sufficient to choose the set $\mathcal{F}$ from the facets
formed by cutting planes of type $\nu^k_1 + \nu^k_2 + \nu^k_3 = 2$ where
$\nu^k_1$, $\nu^k_2$, and $\nu^k_3$ are variables of the CLPD formulation.
Additionally, an adaptive branch \& bound method is suggested in \cite{YaFeWa}
to find better bounds for the minimum distance of a code. On a random
rate-$\frac{1}{4}$ LDPC code with $n=60$, $d_v=3$, $d_c=4$, it is demonstrated
that this yields a better lower bound than the fractional distance does.

\subsection{LP formulations for turbo-like codes}
The various LP formulations outlined so far have in common that they are derived
from a parity-check matrix which defines a
specific code. A different approach is to describe
the encoder by means of a finite state machine, which is the usual way to define
so-called convolutional codes. The bits of the information word are subsequently
fed into the machine, each causing a state change that emits a fixed number
of output bits depending on both the current state and the input.
In a systematic code, the output always contains the input bit.
The codeword, consisting of the concatenation of all outputs,
can thus be partitioned into the systematic part which is a copy of the input
and the remaining bits, being refered to as the parity output.

\begin{figure}\centering

\begin{tikzpicture}[sloped,zero/.style={dashed,->},
                    one/.style={->},
                    vertex/.style={draw,circle,inner sep=.6mm},
                    scale=.8]
\node[vertex] (a0) at (0,0) {$0$};
\node[vertex] (b0) at (1.5,0) {$0$};
\node[vertex] (b2) at (1.5,-2) {$2$};
\draw[zero] (a0) -- node[above] {$0$} (b0);
\draw[one] (a0) -- node[below] {$1$} (b2);
\foreach \pos in {0,...,3} {
  \foreach \state in {0,...,3} {
    \node[vertex] (v\pos\state) at (3 + 1.5*\pos, -\state) {$\state$};
  }
}
\foreach \state in {0,...,3} {
  \coordinate (v4\state) at (3 + 6, -\state);
}
\foreach \pos/\next in {0/1,1/2,2/3} {
  \begin{scope}[zero]
    \draw (v\pos0) -- (v\next0);
    \draw (v\pos1) -- (v\next3);
    \draw (v\pos2) -- (v\next1);
    \draw (v\pos3) -- (v\next2);
  \end{scope}
  \begin{scope}[one]
    \draw (v\pos0) -- (v\next2);
    \draw (v\pos1) -- (v\next0);
    \draw (v\pos2) -- (v\next3);
    \draw (v\pos3) -- (v\next1);
  \end{scope}
}
\begin{scope}[gray]
  \clip (6,0) rectangle (8.3,-3);
  \foreach \pos/\next in {3/4} {
    \begin{scope}[zero]
      \draw (v\pos0) -- (v\next0);
      \draw (v\pos1) -- (v\next3);
      \draw (v\pos2) -- (v\next1);
      \draw (v\pos3) -- (v\next2);
    \end{scope}
    \begin{scope}[one]
      \draw (v\pos0) -- (v\next2);
      \draw (v\pos1) -- (v\next0);
      \draw (v\pos2) -- (v\next3);
      \draw (v\pos3) -- (v\next1);
    \end{scope}
  }
\end{scope}
\draw[zero] (b0) -- node[above] {$0$}(v00);
\draw[one] (b0) -- node[above,near start] {$1$}(v02);
\draw[zero] (b2) -- node[auto] {$0$} (v01);
\draw[one] (b2) -- node[below] {$1$} (v03);
\foreach \pos in {1,2,3} {
  \node[vertex,opacity=0.4,dashed] at (0,-\pos) {$\pos$};
}
\foreach \pos in {1,3} {
  \node[vertex,opacity=0.4,dashed] at (1.5,-\pos) {$\pos$};
}
\end{tikzpicture}
\caption{Excerpt from a trellis graph with four states and initial state
$0$. The style of an edge indicates the according information bit, while the labels refer
to the single parity bit.}\label{fig:trellis}
\end{figure}
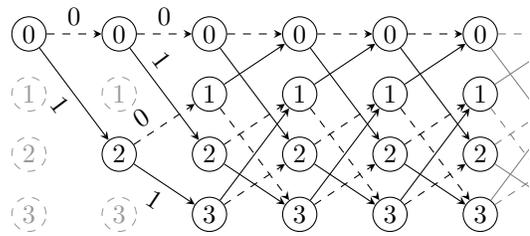

A convolutional code is naturally represented by a
trellis graph (Fig.~\ref{fig:trellis}), which is obtained by
unfolding the state diagram in the time domain. Each vertex of the trellis
represents the state at a specific point in time, while edges
correspond to valid
transitions between two subsequent states and are labelled by the according input and output
bits.
Each path from the starting node to the end node corresponds to a codeword.\footnote{We intentionally
do not discuss trellis termination here and assume that the encoder always ends in a fixed terminal state; cf.~\cite{Shulin} for details.} The
cost of a codeword is derived from the received LLR values and the edge labels
on the path associated with this codeword. See \cite{Shulin} for an in-depth survey of these concepts.

Convolutional codes are the building blocks of turbo codes, which revolutionized
coding theory because of their near Shannon limit error-correcting performance
\cite{Berrou96}. An $(n,k)$ turbo code consists of two convolutional codes $C_a$ and $C_b$,
each of input length $k$,
which are linked by a so-called interleaver that requires the information bits
of $C_a$ to match those of $C_b$ after being scrambled by some permutation $\pi \in \mathbb{S}_k$ which
is fixed for a given code.\footnote{Using exactly two
constituent convolutional encoders eases notation and is the most common case,
albeit not being essential for the concept---in fact, recent development suggest
that the error-correcting performance benefits from adding a third encoder \cite{Berrou2009}.}
It is this coupling of rather weak individual codes and the increase of complexity 
arising therefrom that entails the vast performance gain of turbo codes. A typical turbo code
(and only this case is covered here; it is straightforward to generalize) consists
of two identical systematic encoders of rate $\frac 12$ each. Only one of the 
encoders $C_a$ and $C_b$, however, contributes its systematic part to the resulting codeword,
yielding an overall rate of $\frac 23$, \ie $n=3k$ (since their systematic parts differ only by
a permutation, including both would imply an embedded repetition code).
We thus partition a codeword $x$ into the systematic part $x^s$ and the parity outputs
$x^a$ and $x^b$ of $C_a$ and $C_b$, respectively.

\begin{figure}\centering
\begin{tikzpicture}[font=\scriptsize]
\pgfmathsetseed{6987} % found by experiment, looks nice
\def\maxnode{3} % number of FFG nodes to draw (excluding j)
\def\ffgDistance{1.5cm}
\foreach \component / \y /\placement / \factor in {a/0/above/1, b/-2.5/below/-1} {
  \coordinate(transition\component 0) at (.5,\y);
  \foreach \pos in {1,...,\maxnode} {
    % draw node, connect with its predecessor, and draw parity output edge
    \node[ffgNode] (transition\component\pos) at (\pos*\ffgDistance,\y) {$T^\component_\pos$};
    \pgfmathparse{int(\pos-1)}
    \draw (transition\component \pos) -- node[\placement] {$s^\component_{\pgfmathresult}$} (transition\component\pgfmathresult);
    \draw (transition\component \pos) -- node[left] {$x^\component_\pos$} +($ \factor*(0,.8) $) ;
  }
  \begin{scope}[dashed]
    % draw schematic j-th node
    \draw (transition\component \maxnode) -- ++(1,0) node (end\component) {};
    \node[ffgNode] at ($ (end\component) + (1.5,0) $) (transition\component j) {$T^\component_j$}
      edge node[\placement] {$s^\component_{j}$}  ($ (transition\component j) + (1,0) $) 
	    edge node[\placement] {$s^\component_{j-1}$} ($ (transition\component j) + (-1,0) $);
  \end{scope}
  \draw (transition\component j) -- node[left] {$x^\component_j$}  ($ (transition\component j) + \factor*(0,.8) $);
}
\coordinate(mid) at ($(transitiona0)!.5!(transitionbj)$);
\foreach \pos   in {1,...,\maxnode,j} {
  % draw systematic edges, and complete them to the virtual middle point of the graphic, the x-coordinate
  % being randomly perturbed
  \draw (transitiona\pos) -- node[left] {$x^s_\pos$} ++(0,-.8) coordinate(endpointa\pos)  -- ($ (mid) + (2.5cm*rand,0) $);
  \draw (transitionb\pos) -- node[left] {$x^s_{\pi(\pos)}$} ++(0,.8) coordinate(endpointb\pos) -- ($ (mid) + (2.5cm*rand,0) $);
}
\draw[fill=white,double] ($ (endpointa1) + (-.3,-.15) $) rectangle node[font=\normalfont] {interleaver $\pi$ }($ (endpointbj) +(.3,.15) $);

\end{tikzpicture}
\caption{The factor graph of a turbo code. The interleaver links the systematic bits $x^s$ of both
encoders $C_a$ (upper part) and $C_b$ (lower part).}\label{fig:turboGraph}
\end{figure}
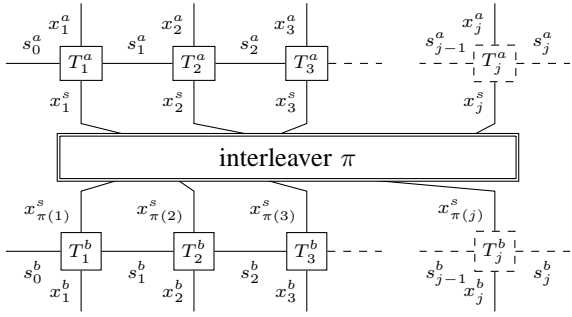

A turbo code can be compactly represented by a so-called Forney-style factor graph (FFG) as shown in Fig.~\ref{fig:turboGraph}. As opposed to Tanner graphs, in an FFG all nodes are functional nodes,
whereas the (half-)edges correspond to variables. In our case, there are variables of two types, namely state
variables $s^\nu_j$ ($\nu \in \{a,b\}$), reflecting the state of $C_\nu$ at time step $j$, and a
variable for each bit of the codeword $x$. Each node $T^\nu_{j}$ represents the indicator function for a
valid state transition in $C_\nu$ at time $j$ and is thus incident to one systematic and one parity
variable as well as the ``before'' and ``after'' state $s^\nu_{j-1}$ and $s^\nu_j$, respectively. Note
that such a node $T^\nu_j$ corresponds to a vertical ``slice'' (often called a segment) of the trellis
graph of $C_\nu$,
and each valid configuration of $T^\nu_j$ is represented by exactly one edge in the respective segment.

Turbo codes are typically decoded by IMPD techniques operating on the factor graph.
Feldman \cite{feld03} in contrast introduced an LP formulation, turbo
code linear programming decoding (TCLPD), for this purpose. This serves as an
example that mathematical programming is a promising 
approach in decoding even beyond formulations based on parity-check matrices.

In TCLPD, the trellis graph of each constituent encoder $C^\nu$ is modeled 
by flow conservation and capacity constraints \cite{Ahuja}, along with
side constraints appropriately connecting the flow variables $f^\nu$ to
auxiliary variables $x^s$ and $x^\nu$, respectively, which embody the codeword bits.

For $\nu \in \{a,b\}$, let $G_\nu = (S_\nu, E_\nu)$ be the trellis according to $C_\nu$,
where $S_\nu$ is the index set of nodes (states) and $E_\nu$ is the set of edges (state transitions)
$e$ in $G_\nu$. Let $s^{\text{start,}\nu}$ and $s^{\text{end,}\nu}$ denote the unique start and
end node, respectively, of $G_\nu$. We can now define a feasible flow $f^\nu$ in the trellis
$G_\nu$ by the system
\begin{gather}  
 \sum_{e \in \operatorname{out}(s^{\text{start,}\nu})}f_e^\nu=1, \label{startstate}
\quad
 \sum_{e \in \operatorname{in}(s^{\text{end,}\nu})}f_e^\nu=1, \\
 \sum_{e \in \operatorname{out}(s)}f_e^\nu = \sum_{e \in \text{in}(s)}f_e^\nu \qquad
\forall\, s \in S_\nu \setminus \{ s^{\text{start,}\nu}, s^{\text{end,}\nu} \},
\label{intermediatestate} \\
f_e^\nu \geq 0 \qquad\forall\, e \in E_\nu.
\end{gather}
Let $I_j^\nu$ and $O_j^\nu$ denote the set of edges in $G_\nu$ whose corresponding input and output bit,
respectively, is a $1$ (both being subsets of the $j$-th segment of $G_\nu$), the following constraints
relate the codeword bits to the flow variables:
\begin{align}
 x^\nu_j = &\sum_{e \in O_j^\nu}f_e^\nu &\text{for }j = 1, \ldots, k\text{ and }\nu \in \{a,b\},
\label{codebitconstraint}\\
 x^s_j = &\sum_{e \in I_j^a}f_e^a &\text{for }j = 1, \ldots, k,
\label{infobitconstraint1} \\
 x^s_{\pi(j)} = &\sum_{e \in I_j^b}f_e^b &\text{for }j = 1, \ldots, k.
 \label{infobitconstraint2}
\end{align}
We can now state TCLPD as 
\begin{align*}   
\min\,& \sum_{\nu \in \{a,b\}} (\lambda^\nu)^T x^\nu + (\lambda^s)^Tx^s\rlap{\quad\quad\text{(TCLPD)}}\\
 \text{s.\,t.}\quad&\text{\eqref{startstate}--\eqref{infobitconstraint2} hold.}
\end{align*}
where $\lambda$ is split in the same way as $x$.

The formulation straightforwardly generalizes to
all sorts of ``turbo-like'' codes, \ie, codes built by convolutional codes plus interleaver conditions.
In particular, Feldman and Karger
have applied TCLPD to repeat-accumulate (RA($l$)) codes \cite{FeldKar}. The
encoder of an RA($l$) repeats the information bits $l$~times, and then sends
them to an interleaver followed by an accumulator, which is a two-state convolutional encoder.
The authors derive bounds on
the error rate of TCLPD for RA codes which were later improved and extended by
Halabi and Even \cite{HalEv} as well as by Goldenberg and Burshtein
\cite{GoldenBursh}.

Note that all $x$ variables in TCLPD are auxiliary: we could replace each occurence
by the sum of flow variables defining it. In doing so,
\eqref{infobitconstraint1} and \eqref{infobitconstraint2} break down to the condition
\begin{equation}
\sum_{e \in I_{\pi(j)}^a} f^a_e = \sum_{e \in I_j^b} f^b_e\qquad\text{for } j=1,\dotsc,k.\label{interleaverconstraints}
\end{equation}
Because the rest of the constraints defines a standard network flow,
TCLPD models a minimum cost flow problem plus the $k$ additional side constraints 
\eqref{interleaverconstraints}. Using a general purpose LP solver does not 
exploit this combinatorial substructure. As was suggested already in
\cite{feld03}, in \cite{Tanatmis2010b} Lagrangian relaxation is applied to \eqref{interleaverconstraints} in order to recover the underlying shortest-path problem.
Additionally, the authors of \cite{Tanatmis2010b} use a  heuristic based on computing the $K$ shortest
paths in a trellis to improve the decoding performance. Via the
parameter $K$ the trade-off between algorithmic complexity and error-correcting
performance can be controlled.

\section{Efficient LP Solvers for BLPD}\label{efflp}  
A successful realization of BLPD requires an efficient LP solver. To this end,
several ideas have been suggested in the literature. CLPD (cf.\
Section~\ref{lprelaxations}) can be considered an efficient LPD approach since
the number of variables and constraints are significantly reduced. We review
several others in this section.

\subsection{Solving the separation problem}\label{efflpA}
The approach of Taghavi and Siegel \cite{TagSie} tackles the large number of
constraints in BLPD2. In their separation approach called adaptive linear
programming decoding (ALPD), not all FS inequalities are included in the LP
formulation as in BLPD2. Instead, they are iteratively added when needed. As in
Definition \ref{defn:SepAlg}, the general idea is to start with a crude LP
formulation and then improve it. Note that this idea can also be used to improve
the error-correcting performance (see Section \ref{incper}).
In the initialization step, the trivial LP $\min \{ \lambda^Tx : x \in [0,1]^n
\}$ is solved. Let $(x^*)^k$ be the optimal solution in iteration $k$. Taghavi
and Siegel show that it can be checked in $O(md_c^{\max}+n\log n)$ time if
$(x^*)^k$ violates any FS inequality derived from $H_{i,.}x=0 \pmod 2$ for all
$i \in I$
(recall that $m\times n$ is the dimension of $H$ and $d_c^{\max}$ is the maximum
maximum check-node degree). This check can be considered as a special case of
the greedy separation algorithm (GSA) introduced in \cite{GroetCard}. If some of
the FS inequalities are violated then these inequalities are added to the
formulation and the modified LP is solved again with the new inequalities. ALPD
stops if the current optimal solution $(x^*)^k$ satisfies all FS inequalities.
If $(x^*)^k$ is integral then it is the ML codeword, otherwise an error is
output. ALPD does not yield an improvement in terms of frame error rate since
the same solutions are found as in the formulations in the previous section.
However, the computational complexity is reduced.

An important algorithmic result of \cite{TagSie} is that ALPD converges to the
same optimal solution as BLPD2 with significantly fewer constraints. It is shown
empirically that in the last iteration of ALPD, less constraints than in the
formulations BLPD2, BLPD3, and CLPD are used. Taghavi and Siegel \cite{TagSie}
prove that their algorithm converges to the optimal solution on the fundamental
polytope after at most $n$~iterations with at most $n(m+2)$~constraints.

Under the binary-input additive white Gaussian noise channel (BIAWGNC),
\cite{TagSie} uses various random $(d_v,d_c)$-regular codes to test the effect
of changing the check node degree, the block length, and the code rate on the
number of FS inequalities generated and the convergence of their algorithm.
Setting $n=360$ and rate $R=\frac{1}{2}$, the authors vary the check node degree
in the range of $4$ to $40$ in their computational testing. It is observed that
the average and the maximum number of FS inequalities remain below $270$. The
effect of changing block length $n$ between $30$ and $1920$ under
$R=\frac{1}{2}$ is demonstrated on a $(3,6)$-regular LDPC code. 
For these codes, it is demonstrated that the number of FS inequalities used in
the final iteration is generally between $0.6n$ and $0.7n$. Moreover, it is
reported that the number of iterations remain below $16$. The authors also
investigate the effect of the rate on the number of FS inequalities created.
Simulations are performed on codes with $n=120$ and $d_v=3$ where the number of
parity checks $m$ vary between $15$ and $90$. For most values of $m$ it is
observed that the average number of FS inequalities ranges between $1.1m$ and
$1.2m$. For ALPD, BLPD2, and SPAD ($50$ iterations), the average decoding time
is testet for $(3,6)$-regular and $(4,8)$-regular LDPC codes with various block
lengths. It is shown that ALPD  outperforms BLPD with respect to computation
time, whil still being slower than SPAD. Furthermore, increasing the check node
degree does not increase the computation time of ALPD as much as the computation
time of BLPD. The behavior of ALPD, in terms of the number of iterations and the
FS inequalities used, under increasing SNR is tested on a $(3,6)$-regular LDPC
code with $n=240$. It is concluded that ALPD performs more iterations and uses
more FS inequalities for the instances it fails. Thus, decoding time decreases
with increasing SNR.

In \cite{Taghavi11Efficient} ALPD is improved further in terms of complexity. The authors
use some structural properties of the fundamental polytope. Let $(x^*)^k$ be an
optimal solution in iteration $k$. In \cite{TagSie} it is shown that, if
$(x^*)^k$ does not satisfy an FS inequality derived from check node $i$, then
$(x^*)^k$ satisfies all other FS inequalities derived from $i$ with strict
inequality. Based on this result, Taghavi \etal \cite{Taghavi11Efficient} modify ALPD and
propose the decoding approach we refer to as modified adaptive linear
programming decoding (MALPD). In the $(k+1)^\text{th}$ iteration of MALPD, it is
checked in $O(m d_c^{\max})$ time if $(x^*)^k$ violates any FS inequality
derived from $H_{i,.}x=0 \pmod 2$ for some $i \in I$. This check is performed
only for those parity checks $i \in I$ which do not induce any active FS
inequality at $(x^*)^k$. Moreover, it is proved that inactive FS inequalities at
iteration $k$ can be dropped. In any iteration of MALPD, there are at most $m$
FS inequalities. However, the dropped inequalities might be inserted again in a
later iteration; therefore the number of iterations for MALPD can be higher than
for ALPD.

\subsection{Message passing-like algorithms}\label{subsec:messagePassing}
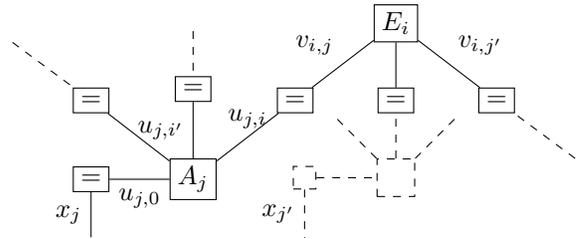
\begin{figure}[htp!]
	\centering
	\begin{tikzpicture}[scale=.8,node distance=.6cm and .8cm]
	 \node[ffgNode] (Ei) at (0,0) {$E_i$};
	 \node[ffgNode,below left=of Ei] (eq1) {$=$} edge node[above left] {$v_{i,j}$} (Ei);
	 \node[ffgNode,below right=of Ei] (eq2) {$=$} edge node[above right] {$v_{i,j'}$}
(Ei);
	 \coordinate[below right=of eq2] edge[dashed] (eq2);
	 \node[ffgNode,below=of Ei] (eqq) {$=$} edge (Ei);
	 \node[ffgNode,dashed,minimum size=5mm,below=of eqq] (inv){} edge[dashed] (eqq)
	 	 edge[dashed] ($ (inv) + (1,1)$)
	 	 edge[dashed] ($ (inv) + (-1,1)$);
	 \node[ffgNode,left=of inv,dashed] (inveq) {} edge[dashed] (inv)
	   edge[dashed] node[left] {$x_{j'}$} ($ (inveq) + (0,-1) $);
	 \node[ffgNode,below left=of eq1] (Aj) {$A_j$} edge node[above] {$u_{j,i}$} (eq1);
	 \node[ffgNode] at ($ (Aj) + (0,1.5) $) (eq3) {$=$} edge node[left] {$u_{j,i'}$}
(Aj);
	 \coordinate[above=of eq3] edge[dashed] (eq3);
	 \node[ffgNode,above left=of Aj] (eq4) {$=$} edge (Aj);
	 \coordinate[above left=of eq4] edge[dashed] (eq4);
	 \node[ffgNode,left=of Aj] (eq5) {$=$} edge node[below]
{$u_{j,0}$} (Aj);
	 \coordinate[below=of eq5] edge node[left] {$x_j$} (eq5);
    \end{tikzpicture}
		\caption{A Forney-style factor graph for PLPD.}
		\label{fig:factorgraph}
	\end{figure}

An approach towards low complexity LPD of LDPC codes was proposed by Vontobel
and Kötter in \cite{VoKo}. Based on an FFG representation of an LDPC code,
they derive an LP, called primal linear
programming decoding (PLPD), which is based on BLPD1. The FFG, shown in
\prettyref{fig:factorgraph}, and the Tanner graph are related as follows.

For each parity check, the FFG exhibits a node $C_i$  which is incident
to a variable-edge $v_{i,j}$ for each $j \in N_i$ and demands those adjacent
variables to form a configuration that is valid for the
local code $C_i$, \ie, their sum must be even. This corresponds to a check node in the Tanner graph and thus
to \eqref{LPD1con3} and \eqref{LPD1con4} except that now there are, for the moment, independent local variables $v_{i,j}$ for each $C_i$.
Additionally, the FFG generalizes the concept of row-wise local codes $C_i$ to
the columns of $H$, in such a way that the $j$\textsuperscript{th} column is considered a local repetition code $A_j$ that requires the auxiliary variables $u_{j,i}$ for each $i \in N_j \cup \{0\}$ to be either all $1$ or all $0$. By this, the variable
nodes of the Tanner graph are replaced by check nodes $A_j$---recall that in an
FFG all nodes have to be check nodes.  There is a third type of factor nodes,
labelled by ``$=$'', which simply require all incident variables to take on the same value. These are used to establish consistency between the row-wise variables $v_{i,j}$ and the column-wise variables $u_{j,i}$ as well as connecting
the codeword variables $x_j$ to the configurations of the $A_j$.

From this discussion it is easily seen that the FFG indeed ensures that
any configuration of the $x_j$ is a valid codeword. The outcome of writing down the constraints for each node and relaxing integrality conditions on all variables is the LP
   \begin{align*}
    &\min  \lambda^Tx \quad\quad \text{(PLPD)} \\
    \text{s.t.}\quad& x_j = u_{j,0} &&j=1,\ldots,n\tk\\
     &u_{j,i} = v_{i,j} &&\forall (i,j) \in I \times J:H_{i,j}=1\tk\\
     &u_{j,i} = \sum_{S \in A_j, S \ni j} \alpha_{j,S} &&\forall i \in N_j, j=1,
\ldots, n\tk\\ 
     &\sum_{S \in A_j} \alpha_{j,S} = 1 &&\text{for all } j=1, \ldots, n\tk\\
     &v_{i,j} = \sum_{S \in E_i, S\ni j} w_{i,S} &&\forall j \in N_i, i=1,
\ldots, m\tk\\ 
     &\sum_{S \in E_i} w_{i,S} = 1 &&\text{for all } i=1, \ldots, m\tk\\ 
     &\alpha_{j,S} \geq 0 &&\forall S \in A_j, j=1, \ldots, n\tk\\
     &w_{i,S} \geq 0 &&\forall S \in E_i, i=1, \ldots, m\text{,}
   \end{align*}
where the sets $E_i$ are defined as in (BLPD1).

While bloating BLPD1 in this manner seems inefficient at first glance, the
reason behind is that the LP dual of PLPD, leads to an FFG which is topologically equivalent to the one of
the primal LP, which allows to use the graphical structure for solving the dual. After manipulating constraints of the dual problem to
obtain a closely related, “softened” dual linear programming decoding (SDLPD)
formulation, the authors propose a
coordinate-ascent-type algorithm resembling the min-sum algorithm and show
convergence under certain assumptions. In this algorithm, all the edges of FFG
are updated according to some schedule. It is shown that the update calculations
required during each iteration can be efficiently performed by the SPAD. The
coordinate-ascent-type algorithm for SDLPD is guaranteed to converge if all the
edges of the FFG are updated cyclically.

Under the BIAWGNC, the authors compare the error-correcting performance of the
coordinate-ascent-type algorithm (max iterations: 64, 256) against the
performance of the MSAD (max iterations: 64, 256) on the $(3,6)$-regular LDPC
code with $n=1000$ and rate $R=\frac{1}{2}$. 
MSAD performs slightly better than the coordinate-ascent-type algorithm. In
summary, Vontobel and Kötter \cite{VoKo} show that it is possible to develop LP based algorithms
with complexities similar to IMPD.

The convergence and the complexity of the coordinate-ascent-type algorithm
proposed in \cite{VoKo} are studied further in \cite{Burshtein} by Burshtein.
His algorithm has a new scheduling scheme and its convergence rate and
computational complexity are analyzed under this scheduling. With this new
scheduling scheme, the decoding algorithm from \cite{VoKo} yields an iterative
approximate LPD algorithm for LDPC codes with complexity in $O(n)$. The main
difference between the two algorithms is the selection and update of edges of
the FFG. In \cite{VoKo} all edges are updated cyclically during one iteration,
whereas in \cite{Burshtein}, only few selected edges are updated during one
particular iteration. The edges are chosen according to the variable values
obtained during previous iterations.

\subsection{Nonlinear programming approach}
As an approximation of BLPD for LDPC codes, Yang \etal \cite{YaFeWa} introduce
the box constraint quadratic programming decoding (BCQPD) whose time complexity
is linear in the code length. BCQPD is a nonlinear programming approach derived
from the Lagrangian relaxation (see \cite{NemWol} for an introduction to
Lagrangian relaxation) of BLPD1. To achieve BCQPD, a subset of the set of the
constraints are incorporated into the objective function. To simplify notation,
Yang \etal rewrite the constraint blocks (\ref{LPD1con3}) and (\ref{LPD1con4}) in
the general form $Ay=b$ by defining a single variable vector $y=(x, w)^T \in \{0,1\}^K$
(so $K$ is the total number of variables in BLPD1) and choosing $A$ and $b$ appropriately.
Likewise, the objective function coefficients are rewritten in a
vector $c$, wich equals $\lambda$ followed by the appropriate number of zeros.
The resulting formulation is $\min \{ c^Ty: Ay=b,
y \in \{0,1\}^K\}$. Using a multiplier $\alpha >0$, the Lagrangian of this
problem is 
\begin{align*}
& \min c^Ty + \alpha(Ay-b)^T(Ay-b) \\
\text{s.t.}\;& 0 \leq y_k \leq 1 \qquad\quad \text{for }k=1, \ldots, K\text{.}
\end{align*} 
If $Ay=b$ is violated then a positive value is added to the original objective
function $c^Ty$, \ie, the solution $y$ is penalized. Setting $Q=2\alpha A^T A$
and $r=c-2\alpha A^T b$ the BCQPD problem 
\begin{align*}
&  \min y^TQy + 2 r^T y {\quad \quad \text{(BCQPD)}}\\
\text{s.t.}\;&  0 \leq y_k \leq 1\qquad\quad \text{for }k=1, \ldots, K 
\end{align*}
is obtained. 
Since $Q$ is a positive semi-definite matrix, \ie, the objective function is
convex, and since the set of constraints constitutes a box, each $y_k$ can be
minimized separately. This leads to efficient serial and parallel decoding
algorithms. Two methods are proposed in \cite{YaFeWa} to solve the BCQPD
problem, the projected successive overrelaxation method (PSORM)  and the
parallel gradient projection method (PGPM). These methods are generalizations of
Gauss-Seidel and Jacobi methods \cite{BertTsi} with the benefit of faster
convergence if proper weight factors are chosen. PSORM and PGPM benefit from the
low-density structure of the underlying parity-check matrix.  
 
One of the disadvantages of IMPD is the difficulty of analyzing the convergence
behavior of such algorithms. Yang \etal showed both theoretically and
empirically that BCQPD converges under some assumptions if PSORM or PGPM is used
to solve the quadratic programming problem. Moreover, the complexity of BCQPD is
smaller than the complexity of SPAD. For numerical tests, the authors use a
product code with block length $4^5=1024$ and rate $(\frac{3}{4})^5=0.237$. The
BIAWGNC is used. It is observed that the PSORM method converges faster than
PGPM. The error-correcting performance of SPAD is poor for product codes due to
their regular structure. For the chosen product code, Yang \etal demonstrate
that PSORM outperforms SPAD in computational complexity as well as in
error-correcting performance.

\subsection{Efficient LPD of SPC product codes}
The class of single parity-check (SPC) product codes is of special interest in
\cite{YaWaFe}. The authors prove that for SPC product codes the fractional
distance is equal to the minimum Hamming distance. Due to this observation, the
minimum distance of SPC product codes can be computed in polynomial time using
FDA. Furthermore, they propose a low complexity algorithm which approximately
computes the CLPD optimum for SPC product codes. This approach is based on the
observation that the parity-check matrix of an SPC product code can be
decomposed into component SPC codes. A Lagrangian relaxation of CLPD is obtained
by keeping the constraints from only one component code in the formulation and
moving all other constraints to the objective function with a penalty vector.
The resulting Lagrangian dual problem is solved by subgradient algorithms (see
\cite{NemWol}). Two alternatives, subgradient decoding (SD) and joint
subgradient decoding (JSD) are proposed. It can be proved that subgradient
decoders converge under certain assumptions. 

The number of iterations performed against the convergence behavior of SD is
tested on the (4,4) SPC product code, which has length $n=256$, rate
$R=\left(\frac34\right)^4\approx 0.32$ and is defined as the product of four SPC
codes of length 4 each. All  variants tested (obtained by keeping the
constraints from component code $j=1,2,3,4$ in the formulation) converge in less
than 20 iterations. For demonstrating the error-correcting performance of SD if
the number of iterations are set to $5,10,20,100$, the (5,2) SPC product code
($n=25$, rate $R=\left(\frac45\right)^2=0.64$) is used. The error-correcting
performance is improved by increasing the number of iterations. Under the
BIAWGNC, this code and the (4,4) SPC product code are used to compare the
error-correcting performance of SD and JSD with the performance of BLPD and MLD.
It should be noted that
for increasing SNR values, the error-correcting performance of BLPD converges to
that of MLD for SPC codes. JSD and SD approach the BLPD curve for the code with
$n=25$. For the SPC product code with $n=256$ the subgradient algorithms perform
worse than BLPD. For both codes, the error-correcting performance of JSD is
superior to SD. Finally, the $(10,3)$ SPC product code with $n=1000$ and rate
$R=(\frac{9}{10})^3 \approx 0.729$ is used to compare the error-correcting
performance of SD and JSD with the SPAD. Again the BIAWGNC is used. It is
observed that SD performs slightly better than the SPAD with a similar
computational complexity. JSD improves the error-correcting performance of the
SD at the cost of increased complexity. 

\subsection{Interior point algorithms}
Efficient LPD approaches based on interior point algorithms are studied by
Vontobel \cite{Vonto08}, Wadayama \cite{WaDa09}, and Taghavi \etal
\cite{Taghavi11Efficient}. The use of interior point algorithms to solve
LP problems as an alternative to the simplex method was initiated by Karmarkar
\cite{Karmakar}. In these algorithms, a starting point in the interior of the
feasible set is chosen. This starting point is iteratively improved by moving
through the interior of the polyhedron in some descent direction until the
optimal solution or an approximation is found. There are various interior point
algorithms and for some, polynomial time convergence can be proved. This is an
advantage over the simplex method which has exponential worst case complexity. 

The proposed interior point algorithms aim at using the special structure of the
LP problem. The resulting running time is a low-degree polynomial function on
the block length. Thus, fast decoding algorithms based on interior point
algorithms may be developed for codes with large block lengths. In particular
affine scaling algorithms \cite{Vonto08}, primal-dual interior point algorithms
\cite{Taghavi11Efficient,Vonto08} and primal path following interior point algorithm
\cite{WaDa09} are considered. The bottleneck operation in interior point methods
is to solve a system of linear equations depending on the current iteration of
the algorithm. Efficient approaches to solve this system of equations are
proposed in \cite{Vonto08,Taghavi11Efficient}, the latter containing an
extensive study, including investigation of appropriate preconditioners for the often ill-conditioned equation system. The speed of convergence to
the optimal vertex of the algorithms in \cite{WaDa09} and \cite{Taghavi11Efficient} under
the BIAWGNC
are demonstrated on a nearly $(3,6)$-regular LDPC code with $n=1008$,
$R=\frac{1}{2}$ and a randomly-generated $(3,6)$-regular LDPC code with
$n=2000$, respectively.

\section{Improving the Error-Correcting Performance of BLPD}\label{incper}
The error-correcting performance of BLPD can be improved by techniques from
integer programming. Most of the improvement techniques can be grouped into
cutting plane or branch \& bound approaches. In this section, we review the
improved LPD approaches mainly with respect to this categorization.

\subsection{Cutting plane approaches}
The fundamental polytope $\mathcal{P}$ can be tightened by cutting plane
approaches. In the following, we refer to valid inequalities as inequalities
satisfied by all points in $\conv(C)$. Valid cuts are valid inequalities which
are violated by some non-integral vertex of the LP relaxation. Feldman \etal
\cite{FeWaKa} already address this concept; besides applying the
``Lift and project'' technique which is a generic tightening method for integer
programs \cite{LovSch}, they also strengthen the relaxation by introducing
redundant
rows into the parity-check matrix (or, equivalently, redundant parity-checks
into the Tanner graph) of the given code (cf.\ Section~\ref{notation}). When
using the BLPD2 formulation, we derive additional FS inequalities from the
redundant parity-checks without increasing the number of variables.
We refer to such inequalities as redundant parity-check (RPC) inequalities.
RPC inequalities may include valid cuts which increase the possibility that LPD
outputs a codeword. An interesting question relates to the types of inequalities
required to describe the codeword polytope $\conv(C)$ exactly. It turns out that
$\conv(C)$ cannot be described completely by using only FS and box inequalities;
the $(7,3,4)$ simplex code (dual of the $(7,4,3)$ Hamming code) is given as a
counter-example in \cite{FeWaKa}. More generally, it can be concluded from
\cite{GrTr2} that these types of inequalities do not suffice to describe all
facets of a simplex code.

RPCs can also be interpreted as dual codewords. As such, for interesting codes
there are exponentially many RPC inequalities. The RPC inequalities cutting off
the non-integral optimal solutions are called RPC cuts \cite{TagSie}.
An analytical study under which circumstances
RPCs can induce cuts is carried out in \cite{VoKoGraph}. Most notably, it is shown that RPCs obtained by adding no more than $\frac{g-2}2$ dual
codewords, where $g$ is the length of a shortest cycle in the Tanner graph, never change the
fundamental polytope.

There are
several heuristic approaches in the LPD literature  to find cut inducing RPCs 
\cite{FeWaKa,MiWaTa,TagSie,TaRuHa}. In \cite{FeWaKa}, RPCs which result from
adding any two rows of $H$ are appended to the original parity-check matrix. The
authors of \cite{TagSie}
find RPCs by randomly choosing cycles in the fractional subgraph of the Tanner
graph, which is 
obtained by choosing only the fractional variable nodes and the check nodes
directly connected to them. They give a theorem which states that every possible
RPC cut must be generated by such a cycle.
Their approach is a heuristic one since the converse of that theorem does not
hold. In \cite{MiWaTa} the column index set corresponding to an optimal LP
solution is sorted. By re-arranging $H$ and bringing it to row echelon form, RPC
cuts are searched. In \cite{TaRuHa}, the parity-check matrix is reformulated
such that unit vectors are obtained in the columns of the parity-check matrix
which correspond to fractional valued bits in the optimal solution of the
current LP. RPC cuts are derived from the rows of the modified parity-check
matrix.

The approaches in \cite{DiGoWa}, \cite{TagSie}, and \cite{TaRuHa} rely on a
noteworthy structural property of the fundamental ploytope. Namely, it can be
shown that no check node of the associated Tanner graph (regardless of the
existence of redundant parity-checks) can
be adjacent to only one non-integral valued variable node.

Feldman \etal \cite{FeWaKa} test the lift and project technique on a random
rate-$\frac{1}{4}$ LDPC code with $n=36$, $d_v=3$ and $d_v=4$ under the BIAWGNC.
Moreover,
a random rate-$\frac{1}{4}$ LDPC code with $n=40$, $d_v=3$, and $d_c=4$ is used
to demonstrate the error-correcting performance of BLPD when the original
parity-check matrix is extended by all those RPCs obtained by adding any two
rows of the original matrix. Both tightening techniques improve the
error-correcting performance of BLPD, though the benefit of the latter
is rather poor, due to the abovementioned condition on cycle lengths.

The idea of tightening the fundamental polytope is usually implemented as a
cutting plane algorithm, \ie, the separation problem is solved (see Definition~\ref{defn:SepAlg} and Section~\ref{efflpA}). In cutting plane algorithms, an LP
is solved which contains only a subset of the constraints of the corresponding
optimization problem. If the optimal LP solution is a codeword then the cutting
plane algorithm terminates and outputs the ML codeword. Otherwise, valid cuts
from a predetermined family of valid inequalities are searched. If some valid
cuts are found, they are added to the LP formulation and the LP is resolved. In
\cite{TagSie,MiWaTa,TaRuHa} the family of valid cuts is FS inequalities derived
from RPCs.

In \cite{MiWaTa} the main motivation for the greedy cutting plane algorithm is
to
improve the fractional distance. This is demonstrated for the $(7,4,3)$
Hamming code, the $(24,12,8)$ Golay code and a $(204,102)$ LDPC code. As a byproduct
under the BSC it is shown on the $(24,12,8)$ Golay code and a $(204,102)$
LDPC code that the
RPC based approach of \cite{MiWaTa} improves the error-correcting performance of
BLPD. 

In the improved LPD approach of \cite{TagSie}, first ALPD (see Section
\ref{efflp}) is applied. If the solution is non-integral, an RPC cut search
algorithm is employed. This algorithm can be briefly outlined as follows:
\begin{enumerate}
 \item Given a non-integral optimal LP solution $x^*$, remove all variable nodes
$j$  for which $x_j^*$ is integral from the Tanner graph.
 \item Find a cycle by randomly walking through the pruned Tanner graph.
 \item Sum up (in $\F_2$) the rows $H$ which correspond to the check nodes in
the cycle.
 \item Check if the resulting RPC introduces a cut.
\end{enumerate}
The improved decoder of \cite{TagSie} performs noticeably better than BLPD and
SPAD. This is shown under the BIAWGNC on $(3,4)$-regular LDPC codes with
$n=32,100,240$.

The cutting plane approach of \cite{TaRuHa} is based on an IP formulation of
MLD, which is referred to as IPD. (Note that this formulation was already mentioned
in \cite{breitbach}.)
Auxiliary variables $z \in \mathbb{Z}^m$ model the binary constraints $Hx=0$ over
$\F_2$ in the real number field $\mathbb{R}^n$.  
\begin{align*}
& \min  \lambda^Tx  \qquad \qquad  \text{(IPD)} \\
\text{s.t.} \; & Hx - 2z = 0 \\
&x \in \left\{ 0,1 \right\}^n,\, z \in \mathbb{Z}^m
\end{align*}        
In \cite{TaRuHa}, the LP relaxation of IPD is the initial LP problem which is
solved by a cutting plane algorithm. Note that the LP relaxation of IPD is not
equivalent to the LP relaxations given in Section \ref{lprelaxations}. In almost
all improved (in the error-correcting performance sense) LPD approaches reviewed
in this article first the BLPD is run. If BLPD fails, some technique to improve
BLPD is used with the goal of finding the ML codeword at the cost of increased
complexity. In contrast, the approach by Tanatmis \etal in \cite{TaRuHa} does
not elaborate on the solution of BLPD, but immediately searches for cuts which
can be derived from arbitrary dual codewords. To this end, the parity-check
matrix is modified and the conditions under which certain RPCs define cuts are
checked. The average number of iterations performed and the average number of
cuts generated in the separation algorithm decoding (SAD) of \cite{TaRuHa} are
presented for the $(3,6)$ random regular codes with $n=40,80,160,200,400$ and
for the $(31,10), (63,39), (127,99), (255,223)$ BCH codes. Both performance
measures seem to be directly proportional to the block length. The
error-correcting performance of SAD is measured on the random regular $(3,4)$
LDPC codes with block length $100$ and $200$, and Tanner's $(155, 64)$ group
structured LDPC code \cite{Tanner2004}. It is demonstrated that the improved LPD
approach of \cite{TaRuHa} performs better than BLPD applied in the adaptive
setting \cite{TagSie} and better than SPAD. One significant numerical result is that SAD
proposed in \cite{TaRuHa} performs much better than BLPD for the $(63,39)$ and
$(127,99)$ BCH codes, which have high-density parity
check matrices. In all numerical simulations the BIAWGNC is used.

Yufit \etal \cite{YuLiBe} improve SAD \cite{TaRuHa} and ALPD \cite{TagSie} by
employing several techniques. The authors propose to improve the
error-correcting performance of these decoding methods by using RPC cuts derived
from alternative parity-check matrices selected from the automorphism group of
$C$, $\Aut(C)$. In the alternative parity-check matrices, the columns of the
original parity-check matrix are permuted according to some scheme. At the first
stage of Algorithm~1 of \cite{YuLiBe}, SAD is used to solve the MLD problem. If the
ML codeword is found then Algorithm~1 terminates, otherwise an alternative
parity-check matrix from $\Aut(C)$ is randomly chosen and the SAD is applied
again. In the worst case this procedure is repeated $N$ times where $N$ denotes
a predetermined constant. A similar approach is also used to improve ALPD in
Algorithm~2 of \cite{YuLiBe}. Yufit \etal enhance Algorithm~1 with two
techniques to improve the error-correcting performance and complexity. The first
technique, called parity-check matrix adaptation, is to alter the parity-check
matrix prior to decoding such that at the columns of the parity-check matrix
which correspond to least reliable bits, \ie, bits with the smallest absolute
LLR values, unit vectors are obtained. The second technique, which is motivated
by MALPD of \cite{Taghavi11Efficient}, is to drop the inactive inequalities at each iteration
of SAD, in order to avoid that the problem size increases from iteration to
iteration. Under the BIAWGNC, it is demonstrated on the $(63,36,11)$ BCH code
and the $(63,39,9)$ BCH code that SAD can be improved both in terms of
error-correcting performance and computational complexity.  

\subsection{Facet guessing approaches}\label{FacGues}
Based on BLPD2, Dimakis \etal \cite{DiGoWa} improve the error-correcting
performance of BLPD with an approach similar to FDA (see Section~\ref{basics}).
They introduce facet guessing algorithms which iteratively solve a sequence of
related LP problems. 
Let $x^*$ be a non-integral optimal solution of BLPD, $x^\text{ML}$ be the ML
codeword, and $\mathcal{F}$ be a set of faces of $\mathcal{P}$ which do not
contain $x^*$. This set $\mathcal{F}$ is given by the set of inequalities which
are not active at $x^*$.

The set of active inequalities of a pseudocodeword $v$ is denoted by
$\mathbb{A}(v)$. In facet guessing algorithms, the objective function
$\lambda^Tx$ is minimized over $f \cap \mathcal{P}$ for all $f \in \mathcal{K}
\subseteq \mathcal{F}$ where $\mathcal{K}$ is an arbitrary subset of
$\mathcal{F}$. The optimal solutions are stored in a list. In random facet
guessing decoding (RFGD), $\left|\mathcal{K}\right|$  of the faces $f \in
\mathcal{F}$ are chosen randomly. If $\mathcal{K}=\mathcal{F}$ then exhaustive
facet guessing decoding (EFGD) is obtained. From the list of optimal solutions,
the facet guessing algorithms output the integer solution with  minimum
objective function value. It is shown that EFGD fails if there exists a
pseudocodeword $v \in f$ such that $\lambda^Tv<\lambda^Tx^\text{ML}$ for all $f
\in \mathbb{A}(x^\text{ML})$. For suitable expander codes this result is
combined with the following structural property of expander-based codes also proven by
the authors. The number of active inequalities at some codeword is much higher
than at a non-integral pseudocodeword. Consequently, theoretical bounds on the
decoding success conditions of the polynomial time algorithms EFGD and RFGD for
expander codes are derived. The numerical experiments are performed under the
BIAWGNC, on Tanner's $(155, 64)$ group-structured LDPC code and on a random LDPC
code with $n=200$, $d_v=3$, $d_c=4$. For these codes the RFG algorithm performs
better than the SPAD. 

\subsection{Branch \& bound approaches}
Linear programming based branch \& bound is an implicit enumeration technique in
which a difficult optimization problem is divided into multiple, but easier
subproblems by fixing the values of certain discrete variables. We refer to
\cite{NemWol} for a detailed description. Several authors improved LPD using the
branch \& bound approach. 

Breitbach \etal \cite{breitbach} solved IPD by a branch \& bound approach.
Depth-first and breadth-first search techniques are suggested for exploring the
search tree. The authors point out the necessity of finding good bounds in the
branch \& bound algorithm and suggest a neighborhood search heuristic as a means
of computing upper bounds.
In the heuristic, a formulation is used which is slightly different to IPD. We
refer to this formulation as alternative integer programming decoding (AIPD).
AIPD can be obtained by using error vectors.
Let $\bar{y}=\frac{1}{2} \left( 1-\textnormal{sign}(\lambda) \right)$ be the
hard decision for the LLR vector $\lambda$ obtained from the BIAWGNC. Comparing 
$\bar{y} \in \{0,1\}^n$ with a codeword $x \in C$ results in an error vector $e
\in \{0,1\}^n$, i.e., $e=\bar{y} + x\pmod 2$. Let $s=H\bar{y}$, and define $\bar
\lambda$ by $\bar\lambda_i = \abs{\lambda_i}$. IPD can be reformulated as
	\begin{align*}
	& \min  \bar\lambda^Te  \quad \quad \text{(AIPD)}  \\
	\text{s.t. }&He-2z=s \\
	& e \in \left\{ 0,1 \right\}^n,\, z \in \mathbb{Z}^m\text{.}
	\end{align*}  
In the neighborhood search heuristic of \cite{breitbach}, first a feasible
starting solution $e^0$ is calculated by setting the coordinates of $e^0$
corresponding to the $n-m$ most reliable bits (\ie,
those $j \in J$ such that $\left|y_j\right|$ are largest) to $0$. These are the
non-basic variables while the $m$ basic variables are found from the vector $s
\in \{0,1\}^m$. Starting from this solution a neighborhood search is performed
by exchanging basic and non-basic variables. The tuple of variables yielding a
locally best improvement in the objective function is selected for iterating to
the next feasible solution.

In \cite{breitbach}, numerical experiments are performed under the BIAWGNC, on
the $(31,21,5)$ BCH code, the $(64,42,8)$ Reed-Muller code, the $(127,85,13)$
BCH code and the $(255,173,23)$ BCH code.
The neighborhood search with single position exchanges
performs very similar to MLD for the $(31,21,5)$ BCH code. As the block length
increases the error-correcting performance of the neighborhood search with single
position exchanges gets worse. An extension of this heuristic allowing two
position exchanges is applied to the $(64,42,8)$ Reed-Muller code, the
$(127,85,13)$ BCH code, and the $(255,173,23)$ BCH code.
The extended neighborhood search heuristic
improves the error-correcting performance at the cost of increased
complexity. A branch \& bound algorithm is simulated on the $(31,21,5)$ BCH code
and different search tree exploration schemes are investigated. The authors
suggest a combination of depth-first and breadth-first search. 

In \cite{DrYeYi}, Draper \etal improve the ALPD approach of \cite{TagSie} with a
branch \& bound technique. Branching is done on the least certain variable,
i.e., $x_j$ such that $\left|x^*_j - 0.5\right|$ is smallest for $j \in J$.
Under the BSC, it is observed on Tanner's $(155,64,20)$ code that the ML
codeword is found after few iterations in many cases. 
    
In \cite{YaFeWa} two branch \& bound approaches for LDPC codes are introduced.
In ordered constant depth decoding (OCDD) and  ordered variable depth decoding
(OVDD), first BLPD1 is solved. If the optimal solution $x^*$ is non-integral, a
subset $\mathcal{T} \subseteq \mathcal{E}$ of the set of all non-integral bits
$\mathcal{E}$ is chosen. Let $g=\left| \mathcal{T} \right|$. The subset
$\mathcal{T}$ is constituted from the least certain bits. The term ``ordered'' in
OCDD and OVDD is
motivated by this construction. It is experimentally shown in \cite{YaFeWa} that
choosing the least certain bits is advantageous in comparison to a random choice
of bits. OVDD is a breadth first branch \& bound algorithm where the depth of
the search tree is restricted to $g$. Since this approach is common in integer
programming, we do not give the details of OVDD and refer to \cite{NemWol}
instead. For OVDD, the number of LPs solved in the worst case is $2^{g+1}-1$. 

In OCDD, $m$-element subsets $\mathcal M$ of $\mathcal{T}$, i.e., $\mathcal{M} \subseteq
\mathcal{T}$ and $m=\left| \mathcal{M} \right|$, are chosen. Let $b \in
\{0,1\}^m$. For any $\mathcal{M} \subseteq \mathcal{T}$, $2^m$ LPs are solved,
each time adding a constraint block 
\[
x_k=b_k \; \text{ for all } k \in \mathcal{M}
\]
to BLPD1, thus fixing $m$ bits. Let $\hat{x}$ be the solution with the minimum
objective function value among the $2^m$ LPs solved. If $\hat{x}$ is integral,
OCDD outputs $\hat{x}$; otherwise another subset $\mathcal{M} \subseteq
\mathcal{T}$ is chosen. Since OCDD exhausts all $m$-element subsets of
$\mathcal{T}$, in the worst case 
$\left( \substack{g \\ m } \right) 2^m+1$ LPs are solved.

The branch \& bound based improved LPD of Yang \etal \cite{YaFeWa} can be
applied to LDPC codes with short block length. For the following numerical
tests, the BIAWGNC is used. Under various settings of $m$ and $g$ it is shown on
a random LDPC code with $n=60$, $R=\frac{1}{4},$ $d_c=4$, and $d_v=3$ that OCDD
has a better error-correcting performance than BLPD and SPAD. Several
simulations are done to analyze the trade-off between complexity and
error-correcting performance of OCDD and OVDD. For the test instances and
parameter settings\footnote{The parameters $m$ and $g$ are chosen such that OVDD and OCDD have
similar worst case complexity.} used in \cite{YaFeWa} it has been observed on
the above-mentioned code that OVDD outperforms OCDD. This behavior is explained
by the observation that OVDD applies the branch \& bound approach on the most
unreliable bits. On a longer random LDPC code with $n=1024$, $R=\frac{1}{4},$
$d_c=4$,  and $d_v=3$, it is demonstrated that the OVDD performs better than BLPD
and SPAD.  

Another improved LPD technique which can be interpreted as a branch \& bound
approach is randomized bit guessing decoding (RBGD) of Dimakis \etal
\cite{DiGoWa}. RBGD is inspired from the special case that all facets chosen by
RFGD (see Section \ref{FacGues}) correspond to constraints of type $x_j \geq 0$
or $x_j \leq 1$. In RBGD, $k=c\log n$ variables, where $c > 0$ is a constant, are
chosen randomly. Because there are $2^k$ different possibile configurations of
these $k$ variables, BLPD2 is run $2^k$ times with associated constraints for
each assignment. The best integer valued solution in terms of the objective
function $\lambda$ is the output of RBGD. Note that by setting $k$ to $c \log
n$, a polynomial complexity in $n$ is ensured. Under the assumption that there
exists a unique ML codeword, exactly one of the $2^k$ bit settings matches the
bit configuration in the ML codeword. Thus, RBGD fails if a non-integral
pseudocodeword with a better objective function value coincides with the ML
codeword in all $k$ components. For some expander codes, the probablilty that
the RBGD finds the ML codeword is given in \cite{DiGoWa}. To find this
probability expression, the authors first prove that, for some expander-based
codes, the number of non-integral components in any pseudocodeword scales linearly
in block length.

Chertkov and Chernyak \cite{ChertChern} apply the loop calculus approach
\cite{chertkov-2006-73}, \cite{chertkov-2006} to improve BLPD. Loop calculus is
an approach from statistical physics and related to cycles in the Tanner graph
representation of a code. In the context of improved LPD, it is used to either
modify objective function coefficients \cite{ChertChern} or to find branching
rules for branch and bound \cite{chert07}. Given a parity-check matrix and a channel
output, linear programming erasure decoding (LPED) \cite{ChertChern} first
solves BLPD. If a codeword is found then the algorithm terminates. If a
non-integral pseudocodeword is found then a so-called critical loop
is searched by employing loop calculus. The indices of the variable nodes along the
critical loop form an index set $\mathcal M \subseteq J$. LPED lowers the objective
function coefficients $\lambda_j$ of the variables $x_j$, $j \in \mathcal M$, by
multiplying $\lambda_j$ with $\epsilon$, where $0 \leq \epsilon < 1$. After
updating the objective function coefficients, BLPD is solved again. If BLPD does
not find a codeword then the selection criterion for the critical loop is
improved. LPED is tested on the list of pseudocodewords found in
\cite{ChertStep} for Tanner's $(155,64,20)$ code. It is demonstrated that LPED
corrects the decoding errors of BLPD for this code. 

In \cite{chert07}, Chertkov combines the loop calculus approach used in LPED
\cite{ChertChern} with RFGD \cite{DiGoWa}. We refer to the combined algorithm as
loop guided guessing decoding (LGGD). LGGD differs from RFGD in the sense that
the constraints chosen are of type $x_j \geq 0$ or $x_j \leq 1$ where $j$ is in
the index set $M$, the index set of the variable nodes in the critical loop.
LGGD starts with solving BLPD. If the optimal solution is non-integral then the
critical loop is found with the loop calculus approach. Next, a variable $x_j$,
$j \in M$, is selected randomly and two partial LPD problems are deduced. These
differ from the original problem by only one equality constraint $x_j=0$ or
$x_j=1$. LGGD chooses the minimum of the objective values of the two
subproblems. If the corresponding pseudocodeword is integral then the algorithm
terminates. Otherwise the equality constraints are dropped, a new $j \in M$
along the critical loop is chosen, and two new subproblems are constructed. If
the set $M$ is exhausted, the selection criterion of the critical loop is
improved. LGGD is very similar to OCDD of \cite{YaFeWa} for the case that
$g=\left|M\right|$ and $m=1$. In LGGD branching is done on the bits in the
critical loop whereas in OCDD branching is done on the least reliable bits.  As
in \cite{ChertChern}, LGGD is tested on the list of pseudocodewords generated in
\cite{ChertStep} for Tanner's $(155,64,20)$ code. It is shown that LGGD improves
BLPD under the BIAWGNC.

SAD of \cite{TaRuHa} is improved in terms of error-correcting performance by a
branch \& bound approach in \cite{YuLiBe}. In Algorithm~3 of \cite{YuLiBe},
first SAD is employed. If the solution is non-integral then a depth-first branch
\& bound is applied. The non-integral valued variable with smallest LLR value is
chosen as the branching variable. Algorithm 3 terminates as soon as the search
tree reaches the maximally allowed depth $D_p$. Under the BIAWGNC, on the
$(63,36,11)$ BCH code and the $(63,39,9)$ BCH code Yufit \etal \cite{YuLiBe}
demonstrate that the decoding performance of Algorithm 3 (enhanced with
parity-check matrix adaptation) approaches MLD.

\section{Conclusion} \label{concl}
In this survey we have shown how the decoding of binary linear block codes
benefits from a wide range of concepts which originate from  mathematical
optimization---mostly linear programming, but also quadratic (nonlinear) and
integer programming, duality theory, branch \& bound methods, Lagrangian
relaxation, network flows, and matroid theory. Bringing together both fields of
research does lead to promising
new algorithmic decoding approaches as well as deeper structural understanding
of linear block codes in general and special classes of codes---like LDPC and
turbo-like codes---in particular. The most important reason for the success of
this connection is the formulation of MLD as the minimization of a linear
function over the codeword polytope $\conv(C)$. We have reviewed a variety of
techniques of how to approximate this polytope, whose description complexity in
general is too large to be computed efficiently.

For further research on LPD of binary linear codes, two general directions can
be distinguished. One is to decrease the algorithmic complexity of LPD towards reducing
the gap between LPD and IMPD, the latter of which still outperforms LPD in
practice. The other direction aims at increasing error-correcting performance,
tightening up to MLD performance. This includes a continued study of RPCs as
well as the characterization of other, non-RPC facet-defining inequalities of
the codeword polytope.

There are other lines of research related to LPD and IMPD which are not covered
in this article. Flanagan \etal \cite{FlanaganNonBinary} have generalized LP
decoding, along with several related concepts, to nonbinary linear codes.
Another possible generalization is to extend to different channel models
\cite{Cohen08Polya}. Connecting two seemingly different decoding approaches,
structural relationship between LPD and IMPD has been
discussed in \cite{VoKo04Relationship}. Moreover, the discovery  that both decoding methods
are closely related to the Bethe free energy approximation, a tool from
statistical physics, has initiated vital research \cite{YedidiaBP}. Also, of
course, research on IMPD itself, independent of LPD, is still ongoing with high
activity. A promising direction of research is certainly the application of
message passing techniques to mathematical programming problems beyond LPD
\cite{Bayati08}.
\begin{table}[htp!]
  \centering
  \caption{List of Abbreviations}
  \begin{tabular}{|l|l|}
  \hline
	AIPD & alternative integer programming decoding\\
	\hline
	ALPD & adaptive linear programming decoding\\
	\hline
	BCQPD & box constrained quadratic programming decoding\\
	\hline
	BIAWGNC & binary input additive white Gaussian noise channel\\
	\hline
	BLPD & bare linear programming decoding\\
	\hline
	BSC & binary symmetric channel\\
	\hline
	CLPD & cascaded linear programming decoding\\
	\hline
	EFGD & exhaustive facet guessing decoding\\
	\hline
	FDA & fractional distance algorithm\\
	\hline
	FFG & Forney style factor graph\\
	\hline
  FS & forbidden set\\
  \hline
  GSA & greedy separation algorithm\\
  \hline
  IMPD & iterative message passing decoding\\
  \hline
  IP & integer programming\\
  \hline
	IPD & integer programming decoding\\
	\hline
	JSD & joint subgradient decoding\\
	\hline
	LLR & log likelihood ratio\\
	\hline
	LDPC & low-density parity-check\\
	\hline
	LGGD & loop guided guessing decoding\\
	\hline
	LP & linear programming\\
	\hline		
	LPD & linear programming decoding\\
	\hline						
	LPED & linear programming erasure decoding\\
	\hline
	MALPD & modified adaptive linear programming decoding\\
  \hline
  ML & maximum likelihood\\
  \hline
  MLD & maximum likelihood decoding\\
  \hline
	MSAD & min-sum algorithm decoding\\
	\hline
	OCDD & ordered constant depth decoding\\
	\hline
  OVDD & ordered variable depth decoding\\
  \hline
  PGPM & parallel gradient projection method\\
  \hline
  PLPD & primal linear programming decoding\\
  \hline
	PSORM & projected successive overrelaxation method\\
	\hline
  RA & repeat accumulate\\
  \hline
  RBGD & randomized bit guessing decoding\\
  \hline
  RFGD & randomized facet guessing decoding\\
  \hline
  RPC & redundant parity-check\\
  \hline
  SAD & separation algorithm decoding\\
  \hline
  SD & subgradient decoding\\
  \hline
  SDLPD & softened dual linear programming decoding\\
  \hline
  SNR & signal-to-noise ratio\\
  \hline
  SPAD & sum-product algorithm decoding\\
  \hline
  SPC & single parity-check\\
  \hline
  TCLPD & turbo code linear programming decoding \\
  \hline
                \end{tabular}
\label{tab:abbr}
\end{table}
\section*{Acknowledgment}
We would like to thank Pascal O.~Vontobel and Frank Kienle for their comments
and suggestions. We also thank the anonymous referees for the helpful
reviews.


\begin{IEEEbiographynophoto}{Michael Helmling}
Michael Helmling received the Diploma in Mathematics from the University of Kaiserslautern, Germany, in 2011.

Since 2011, he has been a Ph.\, D.\ student with the Optimization Research Group, Department of Mathematics, University of Kaiserslautern.
\end{IEEEbiographynophoto}

\begin{IEEEbiographynophoto}{Stefan Ruzika}
Stefan Ruzika received a M.\,S. degree in mathematics from Clemson University, SC, in 2002 and the M.\,S. and Ph.\,D. degrees in mathematics from the University of Kaiserslautern in 2003 and 2007, respectively.

Since 2008 he is assistant professor at the University of Kaiserslautern. His research interests include coding theory, combinatorial optimization and multiple objective programming. 
\end{IEEEbiographynophoto}

\begin{IEEEbiographynophoto}{Akin Tanatmis}
Akin Tanatmis received the B.\,Sc. degree in Industrial Engineering from
Bilkent University, Turkey, in 2002 and the Diploma in Mathematics and Ph.\,D.\ degrees from the University of Kaiserslautern in 2006 and 2011, respectively.
\end{IEEEbiographynophoto}

\end{document}